\documentclass[hidelinks,12pt]{article}
\usepackage[pdftex]{color, graphicx}
\usepackage{amsmath, amsfonts, amssymb, mathrsfs, mathtools}
\usepackage{dcolumn}
\usepackage{natbib}
\usepackage{caption}
\usepackage{listings}
\usepackage{subcaption}
\usepackage{algpseudocode}
\usepackage[]{algorithm}
\usepackage{algorithmicx}
\usepackage{stackengine}
\usepackage{multirow}
\usepackage{hyperref}

\usepackage{diagbox}
\usepackage{dsfont}
\usepackage{bm}

\bibpunct{(}{)}{;}{a}{}{,}
\newcommand{\x}{\mathbf{x}}
\newcommand{\X}{\mathbf{X}}
\newcommand{\Y}{\mathbf{Y}}
\newcommand{\K}{\mathbf{K}}
\newcommand{\kk}{\mathbf{k}}

\newcommand{\U}{\mathbf{U}}

\newcommand{\Q}{\mathbf{Q}}
\newcommand{\W}{\mathbf{W}}
\newcommand{\BB}{\mathbf{B}}
\newcommand{\C}{\mathbf{C}}
\newcommand{\D}{\mathbf{D}}
\newcommand{\E}{\mathbf{E}}

\newcommand{\I}{\mathbb{I}}

\newcommand{\Sig}{\mathbf{\Sigma}}
\algrenewcommand{\algorithmiccomment}[1]{\hskip3em\#\# #1}
\algnewcommand{\LeftComment}[1]{\Statex \#\# #1}

\usepackage{setspace}

\begin{document}

	\title{\vspace{-1cm}
		Large-scale local surrogate modeling of\\stochastic simulation experiments}
	\author{
		D.~Austin Cole\thanks{Corresponding author: \href{mailto:austin.cole8@vt.edu}{\tt austin.cole8@vt.edu}. GlaxoSmithKline, formerly at Virginia Tech}
		\and
		Robert B.~Gramacy\thanks{Department of Statistics, Virginia Tech, Blacksburg, VA}
		\and
		 Mike Ludkovski\thanks{Department of Statistics and Applied Probability,
University of California Santa Barbara}
	}
	\date{}
	\maketitle

	\begin{abstract}
	Gaussian process (GP) surrogate modeling for large computer experiments is
	limited by cubic runtimes, especially with data from stochastic simulations
	with input-dependent noise. A popular workaround to reduce computational
	complexity involves local approximation (e.g., LAGP).  However, LAGP has only
	been vetted in deterministic settings. A recent variation utilizing inducing
	points (LIGP) for additional sparsity improves upon LAGP on the
	speed-vs-accuracy frontier. The authors show that another benefit of LIGP over LAGP is
	that (local) nugget estimation for stochastic responses is more natural,
	especially when designs contain substantial replication as is common when
	attempting to separate signal from noise. Woodbury identities, extended in
	LIGP from inducing points to replicates, afford efficient computation in terms
	of unique design locations only.  This increases the amount of local data (i.e.,
	the neighborhood size) that may be incorporated without additional flops,
	thereby enhancing statistical efficiency. Performance of the authors' LIGP upgrades is
	illustrated on benchmark data and real-world stochastic simulation
	experiments, including an options pricing control framework. Results indicate
	that LIGP provides more accurate prediction and uncertainty quantification for
	varying data dimension and replication strategies versus modern alternatives.
	\end{abstract}
	

\section{Introduction}
\label{sec:intro}
Advancements in high-performance computing (HPC) and techniques like
particle transport and agent-based modeling yield an enormous corpus of
stochastic simulation data.
\citet{baker2022analyzing} provides a review of state-of-the-art
simulation and modeling in such contexts as well as current challenges in this field.  Simulators exist with many applications, including disease/epidemics
\citep{johnson2018phenomenological,hu2017sequential,fadikar2018calibrating}
tumor spread \citep{ozik2019learning}, inventory/supply chain management
\citep{hong:nelson:2006,xie2017heteroscedastic}, ocean circulation
\citep{herbei2014estimating}, and radiation/nuclear safety
\citep{werner2018mcnp6}. In many cases --- in particular those
cited above --- the simulator can exhibit input-dependent-noise, or so-called
heteroskedasticity. When the data is noisy, in addition to the usual nonlinear
dynamics in mean structure, a large and carefully designed simulation campaign
is essential for isolating signal.

Modeling is another crucial component for recognizing signal in noisy data.
Gaussian process (GP) regression is the canonical choice of surrogate model
for simulation experiments because it provides accurate predictions and
uncertainty quantification (UQ). These GP features facilitate many downstream
tasks such as calibration, input sensitivity analysis, Bayesian optimization,
and more. See \cite{Santer2018} or \cite{gramacy2020surrogates} for a review
of GPs and for additional context. However, standard GP inference and
prediction scales poorly to large data sets. GP modeling involves a
multivariate normal (MVN) distribution whose dimension $N$ matches the size of
the training data, i.e., $(\X_N, \Y_N)$ for regression. The quadratic cost of
storage and cubic cost of decomposition of covariance matrices for
determinants and inverses involved in likelihoods limits $N$ to the small
thousands.   For many stochastic computer experiments in modest input
dimension, small simulation campaigns are insufficient for learning whilst
larger $N$ cannot be handled by ordinary GPs.

Numerous strategies abound in diverse literatures (machine learning,
spatial/geostatistics, computer experiments) to scale up GP capabilities in
$N$, with many relying on approximation.  Some take a divide-and-conquer
approach, constructing multiple GPs in segments of a partition of the design
space \citep{Kim2005, Gramacy2008, park2018patchwork}. A local approximate GP
\citep[LAGP;][]{Gramacy2015} offers a kind of infinite-partition, separately
constructing $n \ll N$-sized local neighborhoods around each testing location
$\x'$ in a transductive \citep{vapnik2013nature} fashion.  Small $n$ provides
thrifty processing despite $\mathcal{O}(n^3)$ complexity.   Handling
multiple $\x'$ is parallelizable \citep{gramacy2014massively}.
Predictions are accurate, but a downside to local/partition modeling is
that it furnishes a pathologically discontinuous predictive surface.

Other approaches target matrix decomposition directly by imposing sparsity on
the covariance \citep{titsias2009, aune2014parameter, Wilson2015, gardner2018product,
	Pleiss2018, solin2020hilbert}, the precision matrix \citep{Datta2016,
	katzfuss2021}, or via reduced rank. One rank reduction strategy deploys a set
of $M \ll N$ latent knots, or {\em inducing points}, through which an
$\mathcal{O}(M^2N)$ decomposition is derived \citep[e.g.][]{Williams2001,
	Snelson2006, Banerjee2008, hoffman2013stochastic}. Recently, the inducing
points approach has been hybridized with LAGP \citep[LIGP;][]{cole2021locally}, 
improving computation times and supporting 
larger neighborhood sizes $n$.

GP modeling technology for computer experiments is often tailored to the
no-noise (i.e., deterministic simulation) or constant (iid) noise case, which
also remains true when scaling up via approximation.   Partition-based schemes
are an important exception, as region-based statistically independent modeling
imparts a degree of non-stationarity, potentially in both mean and variance. 
Few mechanisms exist to limit or control partitioning. Computational independence, which speeds up the analysis by allowing multiple
numerical procedures to be carried out in parallel, can be tightly coupled to
regional statistical independence. LAGP is an exception,
in that it allows the user to control compute time directly through the
neighborhood size, $n$. Although the software \citep[\texttt{laGP};][]{gramacy2016lagp}
supports local inference of so-called nugget hyperparameters, its performance for noisy data is often
disappointing because small $n$ may cause noise to be misinterpreted as signal,
%
and exacerbate predictive discontinuity.

Expressly accommodating input-dependent noise has been explored in the
(global/ordinary) GP literature \citep[e.g.,][]{goldberg1997regression,
	kersting2007most}, introducing $N$ latent nugget variables which must be
learned alongside the usual tunable kernel quantities. However, this strategy
is in the opposite direction on the computational efficiency frontier.
Stochastic kriging \citep{ankenman2010} offers thriftier computation for GP
modeling under heteroskedasticity as long as the per-run degree of replication
is high.  This allows latent variables to be replaced with independent
moment-based variance estimates. Heteroskedastic GPs \citep[HetGP;][]{hetGP1}
combine these two ideas: using latent variables for each of $\bar{N}$ unique
input locations and leveraging replication in design, but not requiring a
minimum degree.  The so-called ``Woodbury identities'' \citep{Harville2011}
facilitate exact GP inference and prediction in $\mathcal{O}(\bar{N}^3)$ time,
regardless of $N$. \citet{hetGP2} showed how replication, especially in high
noise regions of the input space, can be essential for both statistical and
computational efficiency. Yet even though $\bar{N} \ll N$, big $\bar{N}$ may
be needed to map out the input space, which can be limiting.

In this paper we propose that, by porting the Woodbury approach from HetGP
over to the LIGP framework, one achieves the best of both worlds.  Relatively
few local-$m$ inducing points could support a much larger number of local
unique inputs $\bar{n}$ representing a much larger neighborhood of $n$ total
observations under replication.  Casting a wide $n$-net is crucial for
separating signal from noise, but would be prohibitive under the original LAGP
regime. In addition to larger neighborhoods, incorporating a state-dependent nugget in the LIGP framework, as is already available in {\tt laGP}, is crucial in estimating noise. Local inducing points, whose number $m \ll \bar{n}
\ll n
\ll N$ may be far fewer than under any of the other sizes, together with a
Woodbury structure for $\bar{n}$ representing the full $n$, make for a
powerful cascade of local information. We derive the quantities involved for
inference for this upgraded LIGP, which include the gradient for numerical
optimization and prediction. Illustrations are provided along the way,
followed by extended benchmarking in both constant noise and heteroskedastic
settings.  We show that this new LIGP capability is both faster and more
accurate than alternatives.

The remaining flow of the paper is as follows. An overview of GP regression
and LIGP is provided in Section \ref{sec:review}. Our contribution, focused on
estimating local noise and incorporating replication in LIGP via Woodbury, is
detailed in Section \ref{sec:gps_with_reps}. Section \ref{sec:results}
summarizes empirical work on synthetic and benchmark examples from
epidemiology and biology. Section \ref{sec:bermuda}
explores this new capability in the context of pricing Bermudan options in computational finance.
Section \ref{sec:discussion} wraps up with a discussion and directions for
future work.

\section{Review}
\label{sec:review}

In this section, we review standard GP surrogate modeling followed by locally induced GPs.

\subsection{Gaussian process regression}
\label{ss:gp_basic}
Consider an unknown function $f: \mathbb{R}^d \rightarrow
\mathbb{R}$ for a set of $d$-dimensional design locations/inputs $\X_N=(\x_1,
\ldots,\x_N)^\top$ and corresponding noisy observations $\Y_N=(y_1, ...,y_N)^\top$ of $f$.
A common surrogate for such data is a Gaussian process (GP), which places an
MVN prior on $f$. The prior is uniquely defined by a mean vector and
covariance (kernel) function $k: \mathbb{R}^d \times \mathbb{R}^d  \rightarrow
\mathbb{R}$. The observation model is
$y(\x_i)=f(\x_i)+\varepsilon_i$, under constant noise, $\varepsilon_i
\stackrel{\mathrm{iid}}{\sim} \mathcal{N}(0, v)$.  Often the iid noise is
hyperparameterized by coupling a nugget $g$ with scale $\tau^2$, as in $v
= \tau^2 g$, so that the full model for responses is $$Y_N\sim \mathcal{N}_N(0,
\tau^2(\K_N + g\I_N)),$$ where
$\I_N$ denotes an $N\times N$ identity matrix. A zero mean simplifies the
exposition without loss of generality.\footnote{Any, even nontrivial, mean
	structure can be subsumed into a covariance kernel if so desired.}  We further assume
zero-centered observations and coded/pre-scaled inputs
\citep{wycoff2021sensitivity} based on inverse distance between rows
of $\X_N$. After pre-scaling, the entries of $\K_N$ can be comprised of kernels $k(\cdot, \cdot)$,
e.g., the so-called isotropic Gaussian kernel:
\begin{equation}
	\label{eq:kernel}
	k_{\theta}(\x_i,\x_j)=\text{exp}\left\{
	-\frac{||\x_i-\x_j||^2}{\theta}\right\}.
\end{equation}
A scalar lengthscale hyperparameter $\theta$ governs the rate of radial
decay of covariance.  Other kernel families such as the Mat\'ern
\citep[][Section 5.3.3]{Stein2012,gramacy2020surrogates}, and so-called
separable forms with lengthscales for each input coordinate are also common.
Our work here is largely agnostic to such choices, as long as $k(\cdot, \cdot)$
is differentiable in the coordinates of $\x$.  Many of the limiting
characteristics of this overly simplistic choice (e.g., infinite smoothness
and isotropy) are relaxed by a local application.

Working with MVNs that involve dense covariance matrices, as is furnished
by $k(\cdot, \cdot)$, entails storage capacity that grows quadratically in
$N$. Prediction and likelihood evaluation for hyperparameter inference, such
as for $(\tau^2, \theta, g)$, involve inverses and determinants requiring
decomposition (usually via Cholesky) that is cubic in $N$.  For brevity, the
relevant formulas are not provided here as they are not needed later.  We
shall provide related expressions momentarily in a more ambitious context.
Special cases of the ones we provide may be found in textbooks alongside
illustrations of many desirable GP properties such as excellent nonlinear
predictive accuracy and UQ.

\subsection{Local approximate GPs}
\label{ss:lagp}

Cubic computational costs are crippling in the modern context of large-$N$
simulation experiments.  As a thrifty workaround, \cite{Gramacy2015}
introduced the local approximate GP (LAGP) which imposes an implicitly sparse
structure by fitting a separate, limited-data GP for each predictive location
$\x'$ of interest. Each fit/prediction is based on an $n \ll N$-sized subset
of data $D_n(\x') = (\X_n(\x'), \Y_n(\x'))$ nearby $\x'$.  Multiple criteria
have been suggested to build this local neighborhood $D_n(\x')$, with
Euclidean nearest neighbor (NN) being the simplest.  Each prediction thus
requires flops in $\mathcal{O}(n^3)$, which can offer dramatic savings and is
embarrassingly parallel for each $\x'$ \citep{gramacy2014massively}.  LAGP was
developed for interpolating deterministic simulations, however the {\tt laGP}
software on CRAN \citep{gramacy2016lagp} includes a nugget-estimating capability for
local smoothing.

\begin{figure*}[ht!]
	\centering
	\includegraphics[trim=0 0 0 0 , clip,width=.98\textwidth]{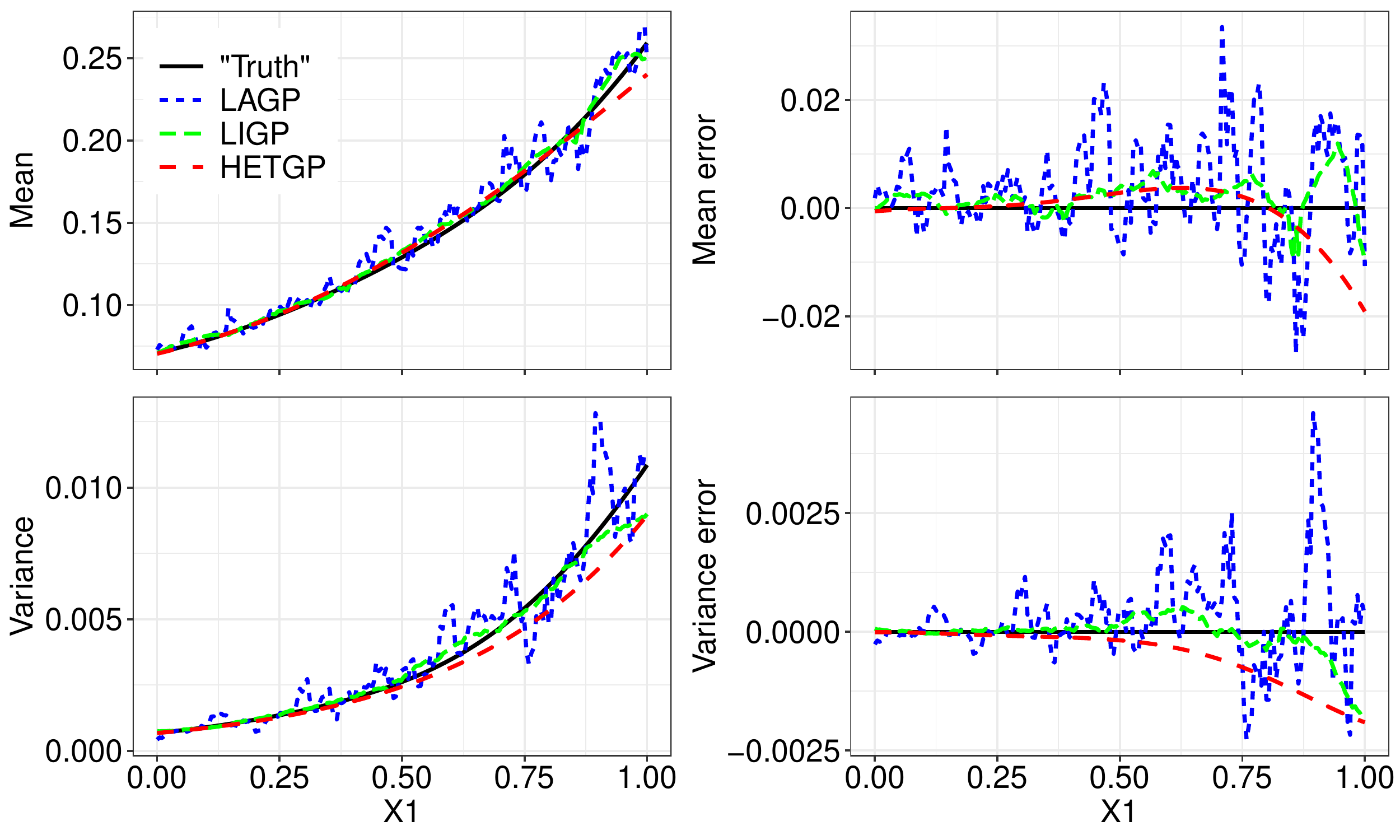}
	\caption{Mean (top) and (bottom) variance estimates (left) and errors (right) for SIR model along the slice $x_2=0.2$.}
	\label{fig:sir_slice}
\end{figure*}

The ``local nugget'' feature underwhelms when applied globally for many $
\x' \in \mathcal{X}$ comprising a dense testing set.
This can be seen first-hand by trying
the examples in the {\tt laGP} documentation.  Figure \ref{fig:sir_slice}
provides a demo in a stochastic simulation context with a
susceptible--infected--recovered (SIR) disease model \citep{hu2017sequential},
implemented as {\tt sirEval} in {\tt hetGP} \citep{hetGP,binois2021hetgp}.  We used a training
set  of $\bar{N}=10000$ unique inputs in $[0,1]^2$, each paired with a
random degree of replication in $\{1,\dots,20\}$, so that the full experiment
consisted of $N \approx 100000$ runs. The views in the figure arise from a
grid of 200 testing locations along the slice $x_2=0.2$. Estimates of the
``true'' mean and variance (for comparison) are based on smoothed averages of
1000 replicates at each location. Comparators include: LAGP.NN with $n=50$,
LIGP.NN with $\bar{n}=100$ ($n\approx 1000$) and $m$=10 inducing points, and
HetGP on a random training subset of size $\bar{N}=1000$. Although each comparator incoroporates a different amount of data, which may seem unfair at face value, this was necessary to balance runtime (to within an order of magnitude, in seconds: 0.5, 6.9, and 15.9 respectively) with accuracy. Observe that
LAGP's estimates (blue dashed lines) of the mean and variance are highly
discontinuous. The error plots in the right column show how LAGP especially
struggles to produce accurate estimates for larger $x_1$.  Details on our
preferred LIGP method (green dashed line), alongside further analysis of this
figure, are the subject of Section \ref{sec:gps_with_reps}.

The problem with applying LAGP to noisy data is both fundamental and
technical. Fundamentally, one needs larger $n$ with noisy data lest the noise
dominate the signal, but this severely cuts into efficiency gains. In a
technical sense, many stochastic simulation experiments involve replication
but the implementation does not adapt its notion of neighborhood accordingly.
In the most extreme case, suppose the nearest input site to $\x'$ has $n$
replicates. Under that configuration LAGP will degenerate without a diversity
of unique training inputs.  Less pathologically, a multitude of nearby
replicates would effectively result in a narrower $n$-neighborhood even though
the number of sufficient statistics is fewer than $n$.  This is inefficient
both statistically and computationally.  Addressing these two issues is the
primary contribution of this paper.

\subsection{Locally induced GPs} \label{ss:ligp}

LIGP \citep{cole2021locally} is a recent extension of the LAGP idea that
relaxes coupling between neighborhood construction and the local covariance
structure.  Rather than imposing a local covariance structure on $Y_n(\x')$
directly via pairwise distances between $\X_n(\x') \subset \X_N$, LIGP uses an
intermediary set of $m$ knots, or so-called {\em inducing points} or {\em
	pseudo inputs} $\boldsymbol{\Psi}_m(\x')$, whose coordinates are not
restricted to $\X_N$.  This yields speedups by reducing the rank of the
covariance structure.  When $m \ll n$ there is a double-speedup,
first through neighborhood structure ($n \ll N$) and then through low rank
local approximation (cubic in $m$ rather than $n$). Conversely, it allows a
much larger neighborhood $n$ without cubic-in-$n$ increase in flops.

Let $\K_m$ denote a kernel matrix built from inducing points $\boldsymbol{\Psi}_m(\x')$ and
$k(\cdot, \cdot)$, e.g., in Eq.~\eqref{eq:kernel}.  We defer a discussion of how
one may choose $\boldsymbol{\Psi}_m(\x')$ to Section \ref{ss:local_neighbor}.
Similarly, write
$\kk_{nm}$ as cross evaluations of the kernel between $\X_n$ and $\boldsymbol{\Psi}_m$.
Adopting the GP approximation with inducing points introduced by
\citet{Snelson2006} yields
\begin{equation} \label{eq:SPGP}
	\Sig_n^{(m)}(\x') = \tau^2\left(\kk_{nm}\K_m^{-1}\kk_{nm}^\top + \Delta_n^{(m)} + g\I_n\right),
\end{equation}
as the MVN covariance deployed for inference/prediction at $\x'$. Here, $\Delta_n^{(m)}=\text{Diag}\{\K_n-\kk_{nm}\K_m^{-1}\kk_{nm}^\top \}$ is a diagonal correction matrix for the covariance and $g\I_n$ expresses the pure noise. Our notation is
suppressing some dependence on $\x'$ to keep the expressions tidy.  Going
forward we shall use $\Omega_n^{(m)}
=\Delta_n^{(m)} + g\I_n$ to express the sum of diagonal matrices in Eq.~\eqref{eq:SPGP},
electing not to embolden $\Omega_n^{(m)}$ like we do for other matrices since
it can be stored as an $n$-vector. When $\boldsymbol{\Psi}_m(\x')\equiv \X_n(\x')$,
$\Sig_n^{(m)}$ reduces to the standard LAGP covariance $\Sig_n(\x') = \tau^2
(\K_n+g\I_n)$ and offers no computational benefit. The cost savings comes
through the decomposition of $\Sig_n^{(m)}$ \eqref{eq:SPGP} through
Woodbury matrix identities. Evaluations of
$\Sig_n^{-1(m)}$ and $\log |\Sig_n^{(m)}|$ may be obtained with
$\mathcal{O}(m^2n)$ flops rather than $\mathcal{O}(n^3)$. Likewise, taking
$\boldsymbol{\Psi}_m = \X_n = \X_N$ with $\Y_n = \Y_N$ yields a global GP without
approximation. 

Hyperparameter inference may be achieved by maximizing the logarithm of the MVN likelihood  $\Y_n \sim
\mathcal{N}(\mathbf{0},\Sig_n^{(m)})$, which may be expressed up to an additive constant as
\begin{align} \label{eq:loglik}
	\ell(D_n(\x'),\boldsymbol{\Psi}_m;\tau^2, \theta, g)
	&\propto  - n\log(\tau^2)-\log |\Q_m^{(n)}|
	+ \log|\K_m| -  \mathbf{1}_n^{\top}\text{log}(\Omega_n^{(m)})\mathbf{1}_n\\ & \qquad -\tau^{-2}\Y_n^\top\left(\Omega^{-1(m)}_n-\Omega_n^{-1(m)}\kk_{nm}\Q_m^{-1(n)}\kk_{nm}^\top\Omega_n^{-1(m)}\right)\Y_n. \nonumber
\end{align}
Above, $\Q_m^{(n)}= \K_m +
\kk_{nm}^{\top}\Omega_n^{-1(m)} \kk_{nm}$
and $\mathbf{1}_n$ is a vector of $n$ ones. Differentiating
Eq.~\eqref{eq:loglik} with respect to $\tau^2$  yields a
closed-form MLE $\hat{\tau}^{2(n,m)}$, see Section
\ref{sec:gps_with_reps}.  A numerical solver like $\tt{optim}$ in
$\sf{R}$ can then work with negative concentrated log-likelihood (i.e., plugging
$\hat{\tau}^{2(n,m)}$ into Eq.~(\ref{eq:loglik})) to estimate $\hat{\theta}^{(n,m)}$
and $\hat{g}^{(n,m)}$.  
It is important to note that \cite{cole2021locally} did not include/estimate a
local nugget $g$, which is  part of our novel contribution. 


For fixed hyperparameters $(\hat{\tau}^2, \hat{\theta}, \hat{g})$, a predictive
distribution for $Y(\x')$ arises as standard MVN conditioning
via an $(n+1)$-dimensional MVN for $(Y(\x'), \Y_n)$.
Using $\kk_m(\x')=k_{\theta}(\boldsymbol{\Psi}_m,\x')$,
the moments of that Gaussian distribution are
\begin{align}
	\label{eq:GPpred}
	\mu_{m,n}(\x')&=\kk_m^{\top}(\x')\Q_m^{-1(n)}\kk_{nm}^\top\Omega_n^{-1(m)}
	\Y_m \hspace{.5cm} \text{and}\\
	\sigma_{m,n}^2(\x')
	& = \hat{\tau}^{2}  \left(k_\theta(\x', \x') + \hat{g} -\kk_m^{\top}(\x')\left(\K_m^{-1}-\Q_m^{-1(n)}\right)\kk_m(\x')\right).\nonumber
\end{align}

Both the log-likelihood (\ref{eq:loglik}), and predictive equations
(\ref{eq:GPpred}) conditional on those values, reduce to LAGP and ordinary GP
counterparts with $\boldsymbol{\Psi}_m = \X_n$ and $\boldsymbol{\Psi}_m = \X_n
= \X_N$ with $\Y_n = \Y_N$, respectively.  Choosing the locations of inducing
points $\boldsymbol{\Psi}_m$, globally or locally, may also be based on the
likelihood \citep{Snelson2006}, however this is fraught with issues
\citep{titsias2009}. \citet{cole2021locally} observed that situating
$\boldsymbol{\Psi}_m$ via classical design criteria like integrated variance,
or so-called $A$-optimal design, led to superior out-of-sample predictive
performance while also avoiding additional cubic calculation.  Specifically
for local analysis in LIGP, they introduced a weighted form of
$\sigma^2_{n,m}(\x')$, centered around $\x'$, and proposed an active learning
scheme that minimized weighted integrated mean-square error (wIMSE) for greedy
selection of the next element $\boldsymbol{\psi}_{m+1}$.  Upgraded details are
provided in Section \ref{sec:gps_with_reps}.
Saving on computation by stylizing the feature of the local designs for
$\boldsymbol{\Psi}_m(\x')$, \citeauthor{cole2021locally}~developed several template schemes that allow to bypass
optimization of wIMSE 
for each $\x'$.

\section{Local smoothing under replication}
\label{sec:gps_with_reps}


LIGP provides an opportunity to reduce compute time while also increasing
neighborhood size, because complexity grows cubically in $m$,  not $\bar{n}$
or $n$.
This is doubly important when there are replicates among $\X_N$.  Without
inducing points, LAGP neighborhoods of size $n$ could have very few unique
inputs $\bar{n} \ll n$.  However, LIGP's Woodbury representation may be extended from inducing points to replicates so that neighborhoods of size $\bar{n}$ may be built, regardless of how much bigger $n$ is, without additional cubic cost.

\subsection{Fast GP inference under replication} \label{ss:gp_infer}
Given a local neighborhood $D_n(\x')=(\X_n(\x'), \Y_n(\x'))$, let
$\bar{\x}_i$, $i=1,\dots,\bar{n}$ represent the $\bar{n} \ll n$ unique
input locations. Notate each unique $\bar{\x}_i$ in $\X_n(\x')$ as having
$a_i\geq 1$ replicates, where $\sum_{i=1}^{\bar{n}}a_i=n$. Let $y_i^{(j)}$
be the response of the $j^\text{th}$ replicate of $\bar{\x}_i$,
$j=1,\dots,a_i$. Without loss of generality, assume the $n$-neighborhood
inputs are ordered so that $\X_n(\x') \equiv \X_n=(\bar{\x}_1,\dots,
\bar{\x}_1, \dots, \bar{\x}_{\bar{n}})^\top$ with each unique input
$\bar{\x}_i$ repeated $a_i$ times, and suppose $\Y_n$ is stacked with responses from
the $a_i$ replicates in the same order. Based on this construction of $D_n$, let
$\X_n=\U\bar{\X}_{\bar{n}}$, where $\U$ is a $n \times \bar{n}$ block
matrix $\U=\text{Diag}(\mathbf{1}_{a_1},\dots,\mathbf{1}_{a_{\bar{n}}})$.
Defining $\U$ in this way, we have $\K_n=\U \K_{\bar{n}}\U^\top$ where
$\K_{\bar{n}} =
(k_\theta(\bar{\x}_i,\bar{\x}_j))_{1\leq,i,j\leq\bar{n}}$,  while
$\U^\top\Omega_n^{(m)}\U=A_{\bar{n}}\Omega^{(m)}_{\bar{n}}$ where
$A_{\bar{n}}=\text{Diag}(a_i,\dots,a_{\bar{n}})$ and
$\Omega_{\bar{n}}^{(m)}
=\text{Diag}\{\K_{\bar{n}}-\kk_{\bar{n}m}\K_m^{-1}\kk_{\bar{n}m}^\top
\} + g\I_{\bar{n}}$.

Using these definitions, we present a new expression for $\Sig_n^{(m)}$,
extending Eq.~\eqref{eq:SPGP} based on the $\bar{n}$ unique input locations
in the local neighborhood:
\begin{equation} \label{eq:SigR}
	\Sig_n^{(m,\bar{n})}
	=\tau^2(\U\kk_{\bar{n}m}\K_m^{-1}\kk_{\bar{n}m}^\top \U^\top + \text{Diag}\{\K_n-\U\kk_{\bar{n}m}\K_m^{-1}\kk_{\bar{n}m}^\top \U^\top \}+g\I_n).
\end{equation}
While $\Sig_n^{(m,\bar{n})}$ in \eqref{eq:SigR} is $n \times n$, we do not
build it in practice. It is an implicit, intermediate quantity which we notate
here as a step toward efficient likelihood and predictive evaluation.

When $m
\ll \bar{n} \ll {n} \ll {N}$, decomposing $\Sig_n^{(m,\bar{n})}$ to obtain
$\Sig_n^{-1(m,\bar{n})} $ and $\log|\Sig_n^{(m,\bar{n})} |$ is much cheaper
than $\mathcal{O}(n^3)$ under the Woodbury identities.  Generically, these are
as follows and may be found in most matrix computation resources
\citep[e.g.,][]{Harville2011}:
\begin{align} \label{eq:woodbury}
	(\BB+\C\D\E)^{-1}&=\BB^{-1}-\BB^{-1}\C(\D^{-1}+\E\BB^{-1}\C)^{-1}\E\BB^{-1}\hspace{.5cm} \text{and}\\
	\log|\BB+\C\D\E|&=\log|\D^{-1}+\E\BB^{-1}\C|+\log|\D|+\log|\BB|, \nonumber
\end{align}
where $\BB$ and $\D$ are invertible matrices of size $n \times n$ and $m \times m$ respectively, and $\C^\top$ and $\E$ are of size $m \times n$.
Efficient decomposition in our local approximation context involves combining LIGP's use of Eq.~(\ref{eq:woodbury}) with inducing points, and \citeauthor{hetGP1}'s (\citeyear{hetGP1}) separate use leveraging replicate structure.
Based on \eqref{eq:SigR}, let $$\BB=\text{Diag}\{\K_n-\U\kk_{\bar{n}m}\K_m^{-1}\kk_{\bar{n}m}^\top \U^\top \}+g\I_n, \quad\D=\K_m^{-1} \quad \text{and }\E=\C^\top=\kk_{\bar{n}m}^\top \U^\top.$$ The result is as follows:
\begin{align} \label{eq:SigiR}
	\Sig_n^{-1(m,\bar{n})}
	=&\tau^{-2}\Big(\Omega_n^{-1(m)}-\U\Omega_{\bar{n}}^{-1(m)}\kk_{\bar{n}m}\Q_m^{-1(\bar{n})}\kk_{\bar{n}m}^{\top}\Omega_{\bar{n}}^{-1(m)}\U^\top \Big) \hspace{.5cm} \text{and}	\\
	\log |\Sig_n^{(m,\bar{n})}|=&\log(\tau^2)+\log |\Q_m^{(\bar{n})}|- \log |\K_m| +\sum_{i=1}^{\bar{n}}a_i\text{log}\omega_i^{(\bar{n}, m)}, \nonumber
\end{align}	
where $\Q_m^{(\bar{n})}= \K_m +
\kk_{\bar{n}m}^{\top}\Lambda_{\bar{n}}^{(m)} \kk_{\bar{n}m}$,
$\Lambda_{\bar{n}}^{(m)}=A_{\bar{n}}\Omega^{-1(m)}_{\bar{n}}$, and
$\omega_i^{(\bar{n}, m)}$ is the $i^{\text{th}}$ diagonal term in
$\Omega^{(m)}_{\bar{n}}$. Eq.~\eqref{eq:SigiR} is key to reducing
computational cost from $\mathcal{O}(nm^2)$ in
Eqs.~(\ref{eq:loglik}--\ref{eq:GPpred}) to $\mathcal{O}(\bar{n}m^2)$. 

The representations in \eqref{eq:SigiR}  allow the log-likelihood
\eqref{eq:loglik} to be expressed as follows
\begin{align} \label{eq:loglikR}
	\ell(D_n,\boldsymbol{\Psi}_m;\tau^2, \theta, g)
	\propto &  - n\log(\tau^2)-\log |\Q_m^{(\bar{n})}| + \log|\K_m| -  \sum_{i=1}^{\bar{n}}a_i\text{log}\omega_i^{(\bar{n}, m)} \\
	&-\tau^{-2}\left(\Y_n^\top \Omega^{-1(m)}_n\Y_n-\bar{\Y}_{\bar{n}}^\top \Lambda_{\bar{n}}^{(m)}\kk_{\bar{n}m}\Q_m^{-1(\bar{n})}\kk_{\bar{n}m}^\top \Lambda_{\bar{n}}^{(m)}\bar{\Y}_{\bar{n}}\right), \nonumber
\end{align}
up to an additive constant.  Above, $\bar{\Y}_{\bar{n}}$ stores the averaged
responses for each unique row $\X_{\bar{n}}$. Notice that when
Eq.~\eqref{eq:SigiR} is applied to the log-likelihood, the $\U$ terms vanish.
In fact, we do not need this quantity or $\Sig_n^{-1(m,\bar{n})}$ except as notational devices. Instead, all matrices in Eq.~\eqref{eq:loglikR} have
dimension in $m$ and/or $\bar{n}$. Although $\Y_n^\top
\Omega^{-1(m)}_n\Y_n$ must still be calculated, this is a product of vectors,
so the entries in $\Omega^{-1(m)}_n$ may be stored and accessed through
$\Omega_{\bar{n}}^{-1(m)}$ and $A_{\bar{n}}$. Their storage and manipulation
are thus linear in $n$.


Differentiating \eqref{eq:loglikR} with respect to $\tau^2$ and solving, gives the MLE:
\begin{equation}\label{eq:nuhatR}
	\hat{\tau}^{2(\bar{n},m)} =  n^{-1}\left(\Y_n^\top\Omega^{-1(m)}_n\Y_n-\bar{\Y}_{\bar{n}}^\top \Lambda_{\bar{n}}^{(m)}\kk_{\bar{n}m}\Q_m^{-1(\bar{n})}\kk_{\bar{n}m}^\top \Lambda_{\bar{n}}^{(m)}\bar{\Y}_{\bar{n}}\right).
\end{equation}
The equation for $\hat{\tau}^{2(\bar{n},m)}$ in \eqref{eq:nuhatR} still relies on all replicates through the full $\Y_n$ and is equivalent to $\hat{\tau}^{2(n,m)}$ as provided by \citeauthor{cole2021locally} (\citeyear{cole2021locally}), Eq. 6.  This part of the calculation is linear in $n$.  However when replicates are present, evaluation in  \eqref{eq:nuhatR} may be much faster owing to the second term being sized in $\bar{n}$ and $m$.
Plugging $\hat{\tau}^{2(\bar{n},m)}$ into Eq.~\eqref{eq:loglikR} yields the
following concentrated negative log likelihood:
\begin{align}
	\label{eq:concentrate_ll}
	-\ell \; (D_n,\boldsymbol{\Psi}_m; \theta, g) &\propto n\log\left(\Y_n^\top\Omega^{-1(m)}_n\Y_n-\bar{\Y}_{\bar{n}}^\top \Lambda_{\bar{n}}^{(m)}\kk_{\bar{n}m}\Q_m^{-1(\bar{n})}\kk_{\bar{n}m}^\top \Lambda_{\bar{n}}^{(m)}\bar{\Y}_{\bar{n}}\right)\\
	&\;\; +\log |\Q_m^{(\bar{n})}| - \log|\K_m| +  \sum_{i=1}^{\bar{n}}a_i\text{log}\omega_i^{(\bar{n}, m)}. \nonumber
\end{align}
The expression above is negated for library-based minimization in search of MLEs $\hat{\theta}^{(\bar{n},m)}$ or $\hat{g}^{(\bar{n},m)}$.  Numerical methods such as BFGS \citep{Byrd1995} work well.  Convergence and computing time of such optimizers is aided by closed-form derivatives, which are tedious to derive but are also available in closed form:
\begin{align} \label{eq:concentrate_deriv}
	-\frac{\partial \ell \; (D_n,\boldsymbol{\Psi}_m, \theta, g)}{\partial \cdot} &\propto n\left(\Y_n^\top\Omega^{-1(m)}_n\Y_n-\bar{\Y}_{\bar{n}}^\top \Lambda_{\bar{n}}^{(m)}\kk_{\bar{n}m}\Q_m^{-1(\bar{n})}\kk_{\bar{n}m}^\top \Lambda_{\bar{n}}^{(m)}\bar{\Y}_{\bar{n}}\right)^{-1}\\
	&\quad\times  \frac{\partial\left(\Y_n^\top\Omega^{-1(m)}_n\Y_n-\bar{\Y}_{\bar{n}}^\top \Lambda_{\bar{n}}^{(m)}\kk_{\bar{n}m}\Q_m^{-1(\bar{n})}\kk_{\bar{n}m}^\top \Lambda_{\bar{n}}^{(m)}\bar{\Y}_{\bar{n}}\right)}{\partial \cdot}\nonumber\\
	&\quad+\text{tr}\left(\Q_m^{-1(\bar{n})}\frac{\partial \Q_m^{(\bar{n})}}{\partial \cdot}\right)- \text{tr}\left(\K_m^{-1}\frac{\partial \K_m}{\partial \cdot}\right) + \sum_{i=1}^{\bar{n}} a_i \frac{\partial \log \omega_i^{(\bar{n}, m)}}{\partial \cdot}. \nonumber
\end{align}
Again, observe that none of these
log-likelihood-derived quantities
(\ref{eq:loglikR}--\ref{eq:concentrate_deriv}) involve matrices sized bigger
than $m\times \bar{n}$.

Conditioning on the estimated hyperparameters leaves us with the following re-expressions of the LIGP predictive equations \eqref{eq:GPpred}:
\begin{align}
	\label{eq:GPpredR}
	\mu_{m,\bar{n}}(\x')
	&=\kk_m^{\top}(\x')\Q_m^{-1(\bar{n})}\kk_{\bar{n}m}^\top\Lambda_{\bar{n}}^{(m)}\bar{\Y}_{\bar{n}} \hspace{.5cm} \text{and}\\
	\sigma_{m,\bar{n}}^2(\x')
	& =\hat{\tau}^{2(\bar{n}, m)}  \left(k_\theta(\x',\x')+\hat{g}^{(\bar{n}, m)} - \kk_m^{\top}(\x')\left(\!\K_m^{-1} - \Q_m^{-1(\bar{n})} \right)\kk_m(\x') \right).\nonumber
\end{align}
Although these look superficially similar to \eqref{eq:GPpred}, the following
remarks are noteworthy. Perhaps the biggest difference
is that the average of replicates
$\bar{\Y}_{\bar{n}}$ is used for the mean.  The Woodbury identity ensures
that the result is the same as
\eqref{eq:GPpred}, both in mean and variance, if the replicate structure is
overlooked.  In both cases, many of the details are buried in matrices whose
$(\bar{n})$ scripts mask a substantial difference in the subroutines that
would be required to build the requisite quantities for \eqref{eq:GPpredR} as
compared to \eqref{eq:GPpred}.

Kriging with pre-averaged responses $\bar{\Y}_{\bar{n}}$, whether locally or
globally, is a common technique for efficiently managing replicates. However,
one must take care to (a) adjust the covariance structure appropriately, and
(b) utilize all replicates in estimating scale $\hat{\tau}^2$.  \citet{hetGP2}
show that $\bar{\Y}_{\bar{n}}$ are not sufficient statistics.  Using
only pre-averaged responses risks
leaving substantial uncertainty un-accounted for in prediction. Our Woodbury
identities provide a full covariance structure (\ref{eq:SigR}), and the
adjustments in $n$-quantities for an appropriate scale estimate
(\ref{eq:nuhatR}). Alternative approaches, such as \citet{ankenman2010}'s
stochastic kriging (SK), also use unique $\bar{n}$ inputs and pre-averaged
outputs.  Asymptotically covering (a), it may be shown that SK yields the best
linear unbiased predictor (BLUP) when conditioning on estimated
hyperparameters. However, the uncertainty in that predictive surface is not
fully quantified.  For example, SK may only furnish confidence intervals on
predictions, not full predictive intervals, because it does not utilize the
full set of sufficient statistics (b). Since our Woodbury application links
Eq.~\eqref{eq:GPpredR} with \eqref{eq:GPpred} identically, a BLUP property
holds (locally) for LIGP without asymptotic arguments, and with full
predictive UQ.

More specifically, consider the following comparison which upgrades
an argument from \cite{hetGP1} to our local inducing points setting.
We show how the MLE of the scale parameter $\tau^2$ is not the same
(conditioned on $\theta, g$) when calculated on the original set of data versus the
unique design locations with averaged responses. Using unique-$\bar{n}$ calculations based on  $\bar{\Y}_{\bar{n}}$ only would result in
$\bar{\tau}^{2(\bar{n},m)}=\bar{n}^{-1}\bar{\Y}_{\bar{n}}^\top \bar{\boldsymbol{\Sigma}}_{\bar{n}}^{-1}\bar{\Y}_{\bar{n}}
$ where
\begin{equation} \label{eq:Sig_avg}
	\bar{\boldsymbol{\Sigma}}_{\bar{n}}^{(m)}=\kk_{\bar{n}m}\K_m^{-1}\kk_{\bar{n}m}^\top+\Delta_{\bar{n}}^{(m)} +gA_{\bar{n}}^{-1},
\end{equation}
weights the nugget (noise) based on the number of replicates $a_i$ at each unique location $\bar{\x}_i$. Our full neighborhood calculation for the local MLE for $\tau^{2}$ with inducing points in Eq.~\eqref{eq:nuhatR} can be rewritten as follows:
\begin{align}\label{eq:tau2_avg}
	\hat{\tau}^{2(\bar{n},m)}
	&=N^{-1}\biggl(\bar{n}\bar{\tau}^{2(\bar{n},m)} + \Y_n^\top\Omega^{-1(m)}_n\Y_n-\bar{\Y}_{\bar{n}}^\top \Lambda_{\bar{n}}^{(m)}\bar{\Y}_{\bar{n}}-\\
	& \qquad\qquad \bar{\Y}_{\bar{n}}^\top \bar{\boldsymbol{\Sigma}}_{\bar{n}}^{-1(m)}((A_{\bar{n}}^{-1}-\I_{\bar{n}})^{-1}\Delta_{\bar{n}}^{-1(m)}+\bar{\boldsymbol{\Sigma}}_{\bar{n}}^{-1(m)})^{-1}\bar{\boldsymbol{\Sigma}}_{\bar{n}}^{-1(m)}\bar{\Y}_{\bar{n}}\biggr). \nonumber
\end{align}
See the Appendix for details. Observe that $\bar{\tau}^{2(\bar{n},m)}$ above in
Eq.~\eqref{eq:Sig_avg} is but a small part of \eqref{eq:tau2_avg}.  The term
following $\bar{\tau}^{2(\bar{n},m)}$ serves as a correction for the variance
estimate.

\subsection{Local neighborhood geography} \label{ss:local_neighbor}

\begin{figure*}[ht!]
	\centering
	\includegraphics[trim=0 10 0 30, clip,width=.65\textwidth]{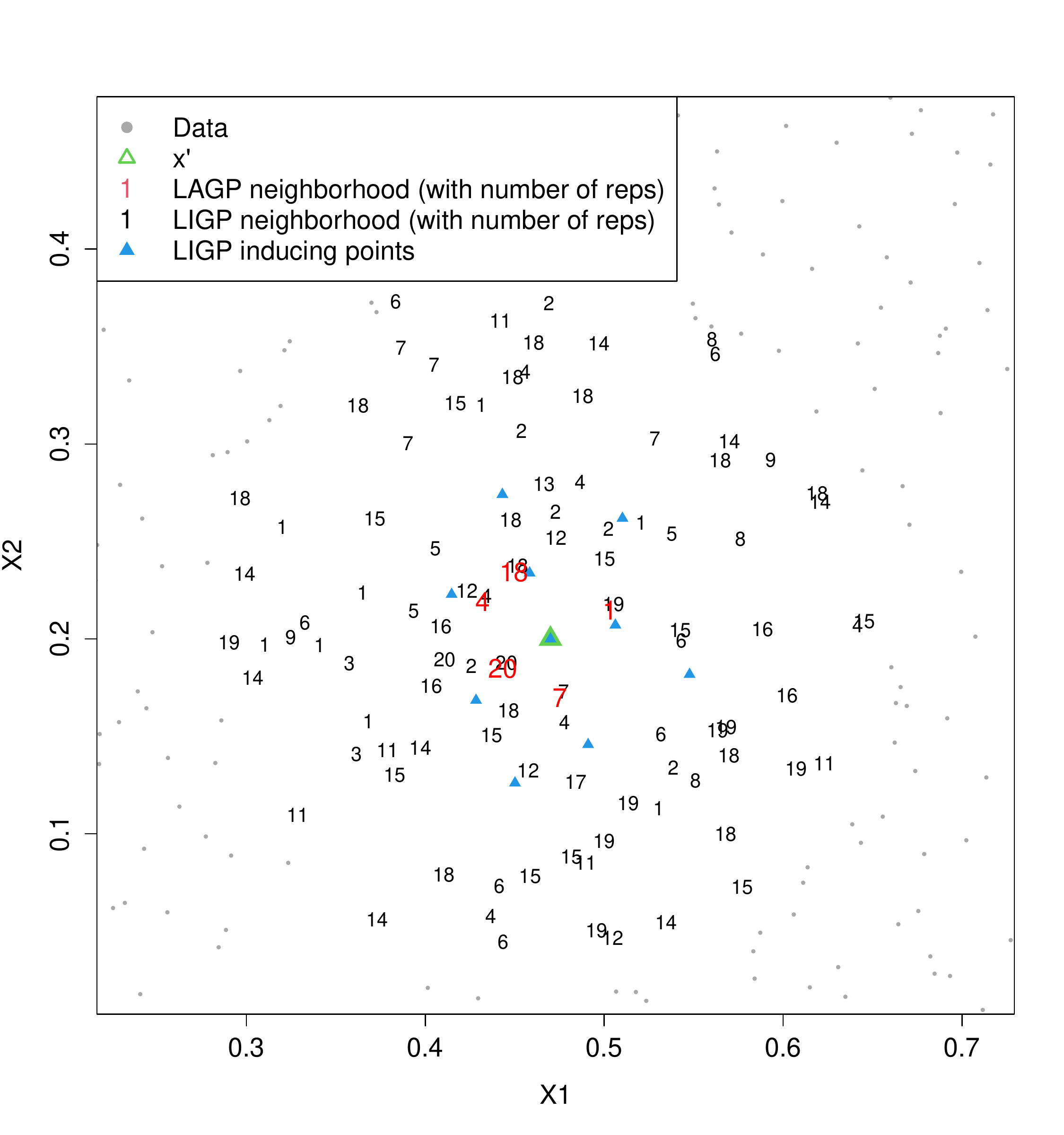}
	\caption{LAGP and LIGP local neighborhoods at $\x'=(0.47, 0.2)$, with the numbers denoting the number of replicates at each location.  Gray dots represent $\X_N \setminus \X_n$.}
	\label{fig:sir_neighborhood}
\end{figure*}

To dig a little deeper into the differences between LAGP and LIGP, especially
regarding handling replicates and LIGP's upgraded capability, consider again
the SIR example introduced in Section \ref{ss:lagp}.  Whereas Figure
\ref{fig:sir_slice} involved a continuum of predictions along a 1d slice,
Figure \ref{fig:sir_neighborhood} shows relevant quantities involved in one of
the locations $\x'=(0.47, 0.2)$ along that slice, indicated by the green
triangle.  Recall that the input space is in 2d with  $\bar{N}=10000$ unique
inputs and replicate degree $a_i\in \{1, 2, \dots, 20\}$ distributed
uniformly.  With such a high degree of replication, LAGP's default $n=50$
neighborhood is small, comprising of only $\bar{n}=5$ unique locations.  As
the green triangle moves along the slice of Figure \ref{fig:sir_slice},
eventually one of those five will be replaced, resulting in a change in
conditioning set of upwards of 1/5 of $\bar{n}$.  This provides some of the
intuition as to why the LAGP surfaces in Figure \ref{fig:sir_slice} are so
``jumpy'' and often inaccurate.

LIGP's implementation of Eqs.~(\ref{eq:loglikR}--\ref{eq:GPpredR}) affords a
much larger amount of data in the local model for commensurate computing
effort through a template of $m=10$ inducing points. These are indicated as
blue triangles in Figure \ref{fig:sir_neighborhood}.   The neighborhood of
$\bar{n}=100$ unique design locations are indicated with their number of
replicates (black numbers), $n=1091$ in total. In this instance, LIGP
conditions upon 20 times more training data than LAGP.  This yields
more accurate and stable mean and variance estimates, shown visually along a
slice in Figure \ref{fig:sir_slice} and summarized in a more expansive Monte
Carlo (MC) exercise in Section \ref{ss:sir}.  Those latter results
additionally show that such accurate predictions can be achieved at a
computational cost that is substantially lower than LAGP.

The location and spacing of the inducing points $\boldsymbol{\Psi}_m(\x')$ in Figure \ref{fig:sir_neighborhood} (blue triangles) come from a local space-filling template
scheme based on wIMSE. The idea is that a local inducing point set of a desired size can be built up greedily, using integrated variance under a Gaussian
measure centered at $\x'$.  An expression using $\bar{n}$ unique
neighborhood elements (together with all replicates, $n$) given \eqref{eq:SigR} is
\begin{align} \label{eq:wimseR}
	&\text{wIMSE}_{\bar{n}}^{(m+1)} (\boldsymbol{\psi}_{m+1}, \x') =\int_{\breve{\x}\in \mathcal{X}}k_\theta(\breve{\x},\x')\frac{\sigma_{m+1,\bar{n}}^2(\breve{\x})}{\tau^2} \; d\breve{\x}  \\
	& \quad =\frac{\sqrt{\theta\pi}}{2}\prod_{k=1}^d\Bigg(\text{erf}\left\{\frac{\x'-a_k}{\sqrt{\theta}}\right\}-\text{erf}\left\{\frac{\x'-b_k}{\sqrt{\theta}}\right\}\Bigg) -\text{tr}\Big\{\Big(\K^{-1}_{m+1}-\Q^{-1(\bar{n})}_{m+1}\Big)\W'_{m+1}\Big\}, \nonumber
\end{align}
where $\text{erf}$ is the Gaussian error function, $a_k,b_k$ are bounds for a hyperrectangle study region
$\mathcal{X} = [a_k, b_k]_{k=1}^d$, and $\W'_{m+1}=\prod_{k=1}^d
\W'_{m+1,k}$.
Again, the resemblance to a similar expression from \citet{cole2021locally}
masks a substantial enhancement in its component parts due to a
double-Woodbury application, as notated by $\bar{n}$ scripts.  The
$(i,j)^\text{th}$ entry of $\W'_{m+1,k}$ is $w_{m+1,k}(\boldsymbol{\psi}_{i},
\boldsymbol{\psi}_{j})$ defined in Eq.~(11) in \citet{cole2021locally}.
A closed-form gradient with respect to $\boldsymbol{\psi}_{m+1}$ may also be
derived.  This is similar in form to \citeauthor{cole2021locally}, so we do
not duplicate it here. Given $\Q^{-1(\bar{n})}_{m+1} \equiv
\Q^{-1(n)}_{m+1}$, the calculation in \eqref{eq:wimseR} and derivative are
equivalent to Eqs.~(10--12) in that paper.  Consequently, a wIMSE optimized
design of inducing points is not affected by the local distribution of
replicates. Under replication $\Q^{-1(\bar{n})}_{m+1}$ can, of course, be
calculated much faster than $\Q^{-1(n)}_{m+1}$, so the optimization is
speedier.

Regularity in local inducing point design also means that simple space-filling
designs can serve as thrifty substitutes to the wIMSE design with similar
predictive accuracy. The qNorm method \citep{cole2021locally} transforms
a space-filling design with the inverse-CDF of a Normal distribution to mimic
an wIMSE inducing point design. This is what was used for Figures
\ref{fig:sir_slice}--\ref{fig:sir_neighborhood}.  Once a set of inducing
points $\boldsymbol \Psi_m(\x')$ is calculated for one $\x'$,
\citet{cole2021locally} show that it can serve as a {\em template} for
predictions at other members of a testing set $\mathcal{X}$ through
displacement and warping, so long as the characteristics of the original
design $\X_N$ are uniform.  Templates based on wIMSE tend to fill the
neighborhood in a more uniform manner than via qNorm, providing a higher
density of inducing points near the predictive location.  This often leads to
more accurate prediction. Yet the reduced time required to build qNorm templates is
also beneficial. We contrast both options in our empirical work in
Section \ref{sec:results}.


\section{Implementation and benchmarking}
\label{sec:results}

Now we provide practical details and report on experimental results showcasing LIGP's potential for superior accuracy and UQ at lower
computational costs against LAGP and HetGP. Our examples include data from a toy
function and real stochastic simulators. All analysis was performed on
an eight-core hyperthreaded Intel i9-9900K CPU at 3.60 GHz. Every effort
was made to take advantage of that distributed resource to minimize compute times.

\subsection{Implementation details}
\label{ss:implement}
Open source implementation of LIGP can be found in the {\tt liGP} package on
CRAN \citep{liGP}. {\sf R} code \citep{R} supporting all examples reported
here and throughout the paper, may be found on our Git repository.
\begin{center}
	\url{https://bitbucket.org/gramacylab/ligp/src/master/noise}
\end{center}
Our main competitors are {\tt laGP} \citep{gramacy2016lagp} and {\tt hetGP}
\citep{hetGP}. While {\tt laGP} is coded in {\sf C} with {\tt OpenMP} for
symmetric multiprocessing (SMC) parallelization ({\sf R} serving only as
wrapper), and {\tt hetGP} leverages substantial {\tt RCpp} \citep{Rcpp}, our
LIGP implementation coded purely in {\sf R} using {\tt foreach} for SMC
distribution. With {\tt laGP} we use the default neighborhood size of $n=50$
built by NN and Active Learning Cohn \citep[ALC;][]{cohn1994adavances},
representing distance- and variance-based (similar to wIMSE)
approaches, respectively. For {\tt hetGP}, we reduce the training data to a
random subset of $\bar{N}=1000$ unique inputs (retaining all replicates) to
keep decompositions tractable. Despite compiled {\sf C/C++} libraries and
thrifty (local/global) data subset choices, our (more accurate and
larger-neighborhood) LIGP models are competitive, time-wise, and sometimes
notably faster.

Our LIGP implementation uses an isotropic Gaussian kernel with scalar
lengthscale $\theta$. To improve numerical conditioning of matrices $\K_m$ and
$\Q_m^{(n)}$ for stable inversion, we augment their diagonals with
$\epsilon_K=10^{-8}$ and $\epsilon_Q=10^{-5}$ jitter \citep{Neal1998},
respectively. Our implementation in {\tt liGP} automatically increases these
values for a particular local model if needed. LAGP
and HetGP results also utilize an isotropic Gaussian kernel. Using separable
local formulations do not significantly improve predictive performance,
especially after globally prescaling the inputs \citep{sun2019emulating}.
Pre-scaling or warping of inputs \citep{wycoff2021sensitivity} has become a
popular means of boosting predictive performance of sparse GPs
\citep[e.g.,][]{katzfuss2020scaled}. In our exercises, we divide by
square-rooted separable global lengthscales obtained from a GP fit to random
$\bar{N}=1000$ data subsets with averaged responses $\bar{\Y}_{\bar{N}}$. See
\cite{gramacy2020surrogates}, Section 9.3.4, for details. The time required
for this, which is negligible  relative to other timings, is not included in
our summaries.

For inducing point designs $\boldsymbol{\Psi}_m(\x')$, we use wIMSE templates
based on Eq.~\eqref{eq:wimseR} built at the center of the design space
$\check{\x}$, determined as the median of $\X_N$ in each dimension. For
initial local lengthscale $\theta^{(0)}$, we adopt a strategy from {\tt laGP}
via the 10\% quantile of squared pairwise distances between members of
$\X_n$.\footnote{In {\tt laGP}, the function providing $\theta^{(0)}$ in this
	way is {\tt darg}.} When selecting each $\boldsymbol{\psi}_{m+1}$ to augment $\boldsymbol{\Psi}_m$,
optimizing wIMSE is conducted via a 20-point multi-start derivative-based L-BFGS-B
\citep{Byrd1995} scheme (using {\tt optim} in {\sf R} to a tolerance of 0.01) peppered within the
bounding box surrounding the neighborhood $\X_n$.  In
addition to the wIMSE design for the inducing point template, we consider
so-called ``qNorm''  templates derived from space-filling designs. These originate
from $m-1$ point Latin hypercube samples \citep[LHS;][]{Mckay:1979} on
the hyperrectangle enclosing $\X_n(\check{\x})$,  augmented with $\check{\x}$
as the $m^\mathrm{th}$ point. These LHSs are warped with the inverse
Gaussian CDF to closely mimic the wIMSE design. 

Individual local neighborhoods and
predictions are made for a set of $N'$ testing locations $\x' \in
\mathcal{X}$ given training data $\{\X_N,\Y_N\}$, neighborhood size $\bar{n}$,
and number of inducing points $m$. We use $\bar{n}=100$ for each experiment,
with $m=10$ and $30$ for the 2d and 4d problems, respectively. Each location $\x'_i$,
for $i = 1, \dots, N'$ proceeds in parallel via 16 {\tt foreach}
threads.\footnote{Two per hyperthreaded core.} 
To estimate scale and lengthscale, we used
Eqs.~(\ref{eq:nuhatR}--\ref{eq:concentrate_deriv}) via {\tt optim} through the
local neighborhoods of $\x'$. To aid in discerning signal from noise
during hyperparameter optimization, we incorporate default priors for $\theta$
and $g$, from {\tt laGP}, described in Appendix A of \cite{gramacy2016lagp}.
Predictions follow Eq.~\eqref{eq:GPpredR}.


\subsection{Benchmark non-stationary data}
\label{ss:syn_results}
The toy 2d function known as Herbie's tooth \citep{herbtooth} is attractive as
a benchmark problem due to its non-stationary mean surface with multiple local
minima. The mean function is defined by $f(x_1, x_2)= -w(x_1)w(x_2)$, $x_1,x_2
\in [-2,2]$, where
$$w(x)=\text{exp}\left\{-(x-1)^2\right\}+\text{exp}\left\{-0.8(x+1)^2\right\}-0.05\sin\left(8(x+0.1)\right).$$
For this experiment we introduce constant noise $\epsilon \sim  N(0,0.02^2)$,
creating a response $y(x_1, x_2) = f(x_1, x_2) + \epsilon$. Each of the 30 MC
repetitions is comprised of a fresh training set of $\bar{N}=10000$ LHS
locations, each with $a_i \stackrel{\mathrm{iid}}{\sim}
\mathrm{Unif}\{1,2,\dots, 20\}$ for $i=1,\dots,\bar{N}$.  Although our later
experiments use a more common, fixed-degree setup for replicates, we choose random $a_i$ here to underscore that LIGP does not require a minimum amount of
replication. A separate LHS of $N'=10000$ out-of-sample testing locations is created
for each MC run. Our LAGP fits include $n=100$ via NN for a slightly fairer
comparison to LIGP, in addition to the other LAGP options described earlier.
\begin{figure*}[ht!]
	\centering
	\includegraphics[trim=0 0 20 20 , clip,width=.49\textwidth]{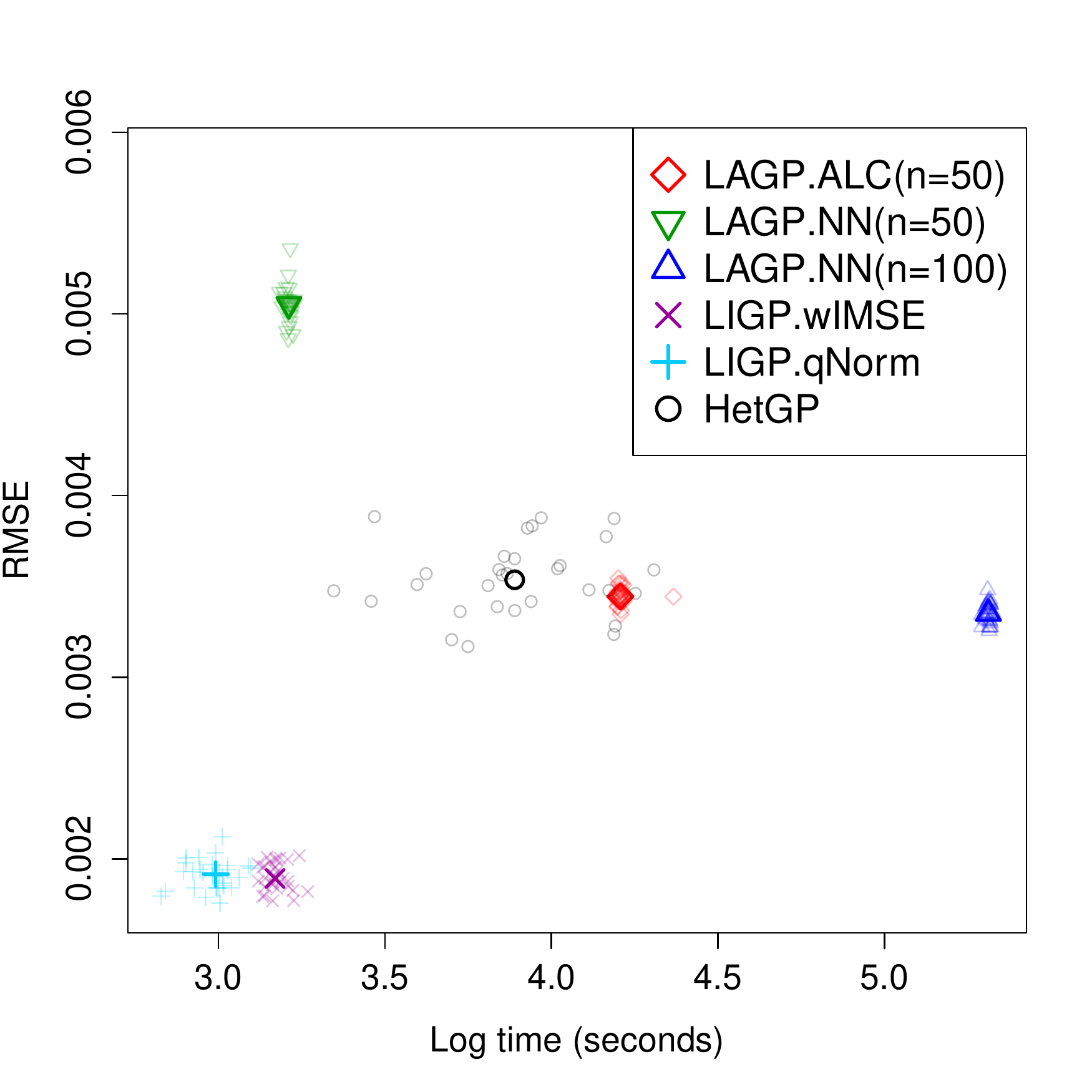}
	\includegraphics[trim=0 0 20 20 , clip,width=.49\textwidth]{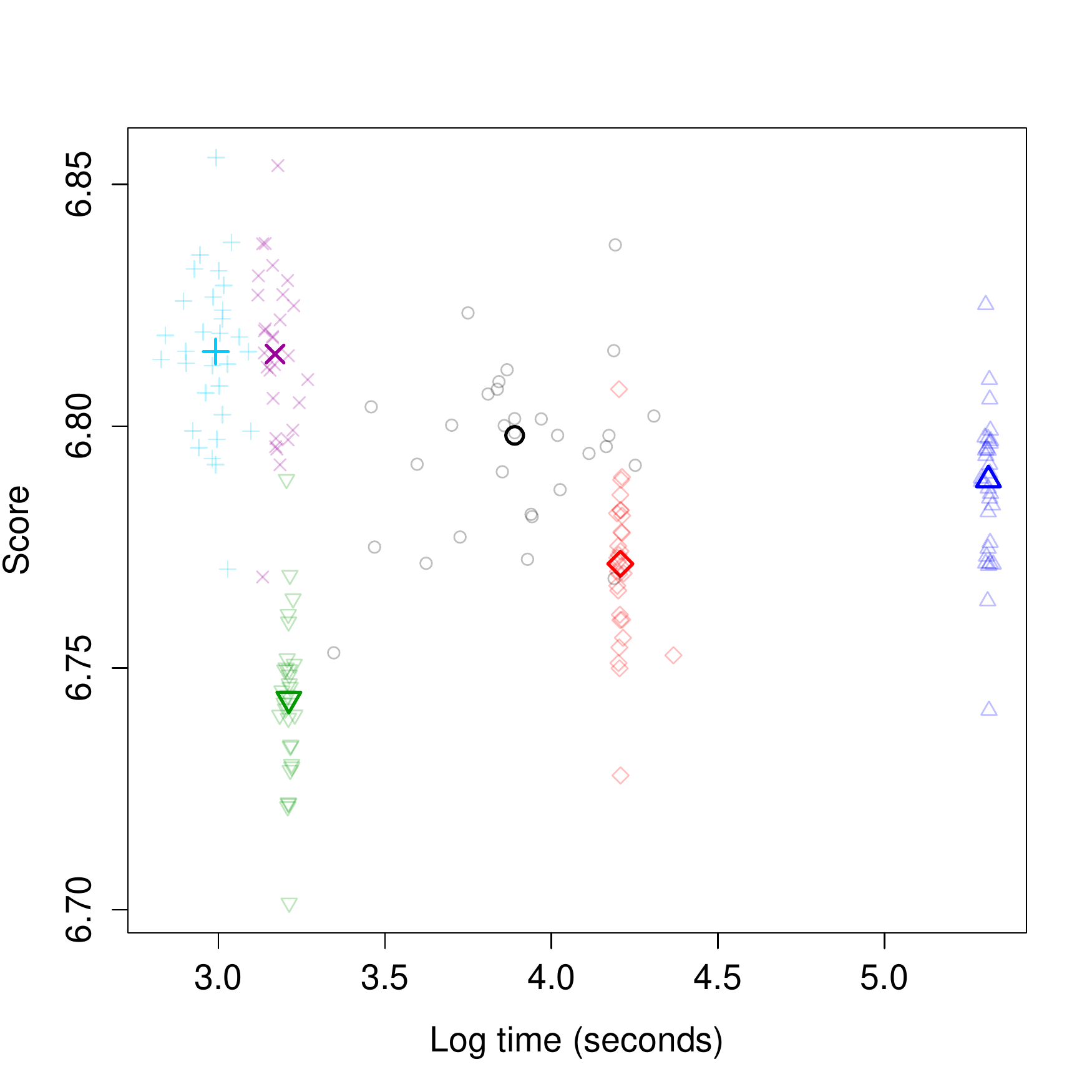}
	\caption{RMSE (left) and score (right) vs.~log compute time over 30 MC
		repetitions for Herbie's tooth experiment. Median statistics are denoted
		with bold markers.}
	\label{fig:ht_results}
\end{figure*}

Figure \ref{fig:ht_results} shows out-of-sample accuracy statistics versus the
logarithm of compute time across the MC repetitions. A particular method's
median statistic, taken marginally across both axes, is represented with a
bold symbol. The left panel shows the root mean-squared prediction error
(RMSE; lower is better) against the (de-noised) truth along the vertical axis.
In the right panel, the vertical axis shows proper scores (higher is better),
combining mean and variance (UQ) accuracy
\citep[][Eq.~(27)]{gneiting2007strictly} against the (noisy) test data. Concretely,
RMSE (lower is better) and score (higher) follow:
\begin{equation*} \label{eq:rmse_score}
	\text{RMSE}=\sqrt{\sum_{i=1}^{N'}(\hat{\mu}(\x_i)-y(\x_i))^2} \hspace{.4cm} \text{and} \hspace{.4cm} \text{Score}=-\sum_{i=1}^{N'}\frac{\left(\hat{\mu}(\x_i)-y(\x_i)\right)^2}{\hat{\sigma}^2(\x_i)}-\sum_{i=1}^{N'}\hat{\sigma}^2(\x_i).
\end{equation*}
LIGP methods yield the fastest predictions with the lowest RMSEs and highest
scores. Our LIGP.qNorm (blue $+$'s) has the quickest compute time. LAGP.NN
with $n=50$ (green inverted triangles) is the fastest among the LAGP models,
but is least accurate. HetGP (black circles) is fit with substantial
sub-setting, which explains its high-variance metric on both time and
accuracy. Notice that LAGP.ALC (red diamonds) and LAGP.NN ($n=100$; navy
triangles) perform similarly for RMSE, but the latter has a higher score. This
supports our claim that a wider net of local training data is needed to
better estimate noise.

\subsection{Susceptible-infected-recovered (SIR) epidemic model}
\label{ss:sir}
We return to the SIR model \citep{hu2017sequential} first mentioned in Section
\ref{ss:lagp}. SIR models are  commonly used for cost-benefit analysis of
public health measures to combat the spread of communicable diseases such as
influenza or Ebola. We look at a simple model with two inputs: $x_1$ the
initial number susceptible individuals; and $x_2$ the number of infected
individuals. In {\tt hetGP} \citep{hetGP}, the function {\tt sirEval} accepts
these two inputs on the unit scale. The response of the model is the expected
aggregate number of infected-days across Markov chain trajectories until the
epidemic ends. The signal-to-noise ratio varies drastically throughout
$\mathcal{X}$, making this model an ideal test problem for LIGP.

\begin{figure*}[ht!]
	\centering
	\begin{subfigure}{.49\textwidth}
		\includegraphics[trim=0 0 20 20 , clip,width=\textwidth]{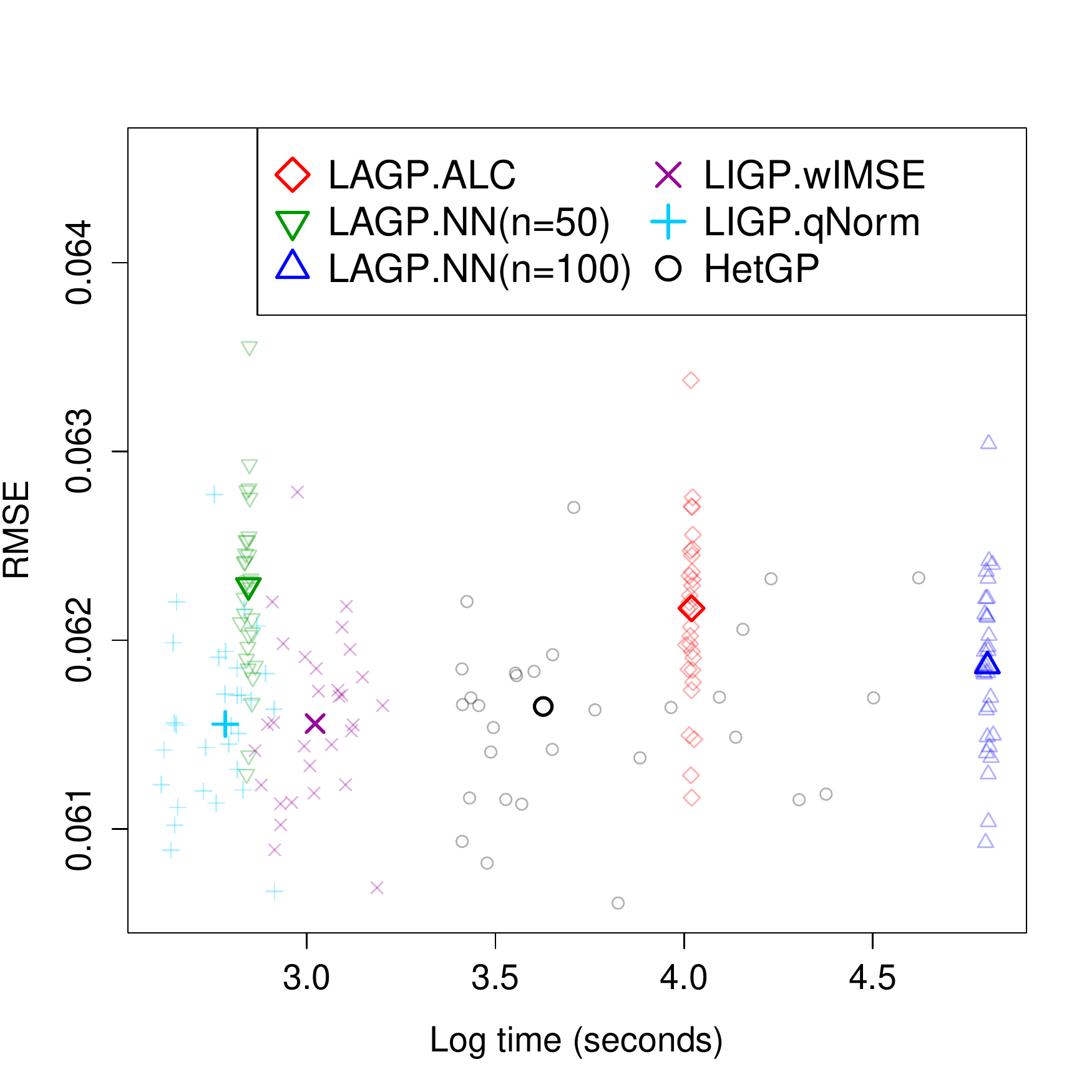}
	\end{subfigure}
	\begin{subfigure}{.49\textwidth}
		\includegraphics[trim=0 0 20 20 , clip,width=\textwidth]{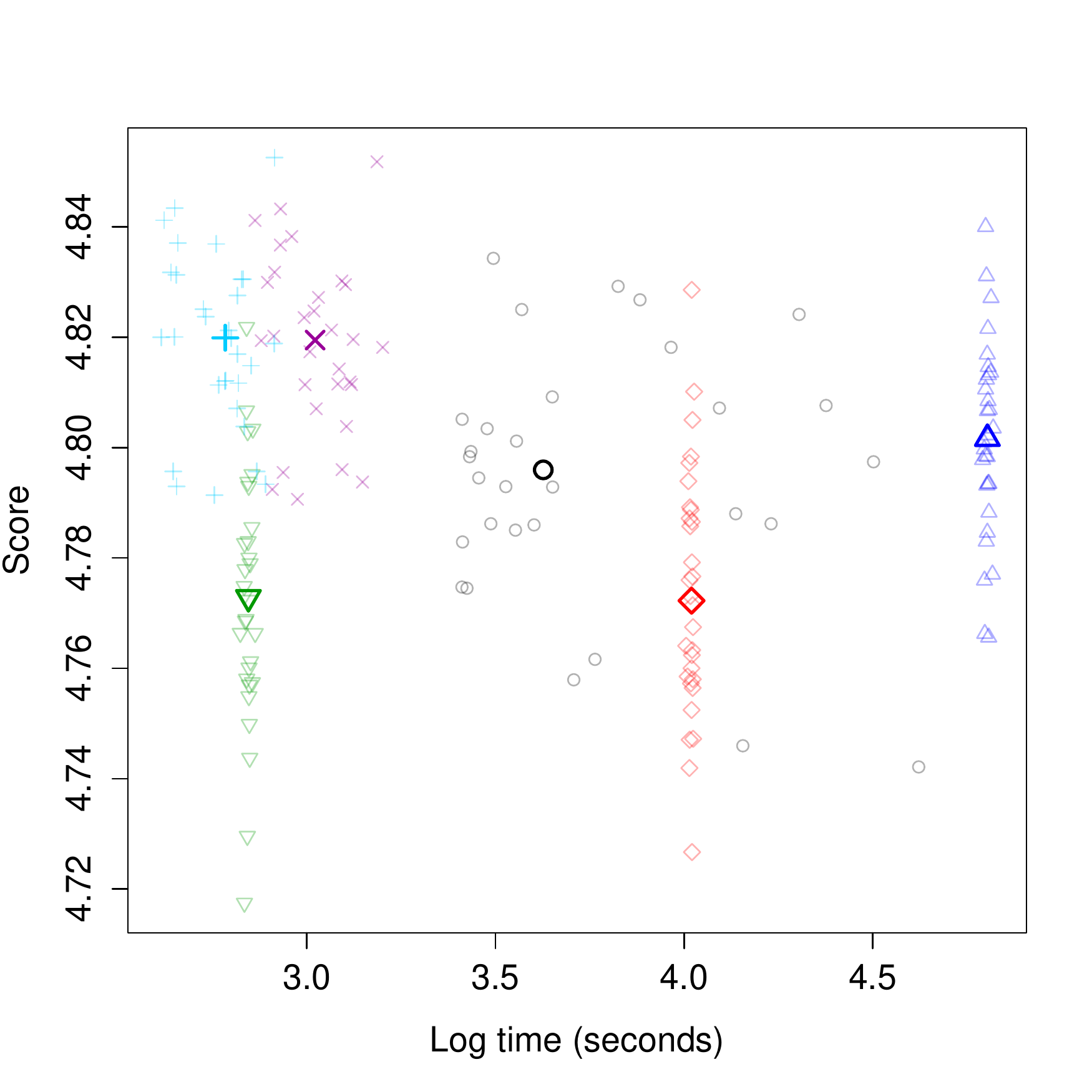}	
	\end{subfigure}
	\footnotesize
	\vspace{0.25cm}
	\begin{tabular}{|l r r|| l r r r|}
		Model & RMSE & $p$-value & Model & Score & $p$-value & Time\\ \hline
		LIGP.qNorm & 0.06156 & 0.0041 & LIGP.qNorm & 4.82 & 0.0031 & 16.2 \\
		LIGP.wIMSE & 0.06156 & 0.6348 & LIGP.wIMSE & 4.819 & $<10^{-9}$ & 20.5 \\
		HetGP & 0.06165 & $<10^{-8}$ & LAGP.NN ($n$=100) & 4.802 & 0.0288 & 122 \\
		LAGP.NN ($n$=100) & 0.06186 & $<10^{-9}$ & HetGP & 4.796 & $<10^{-5}$ & 37.6 \\
		LAGP.ALC ($n$=50) & 0.06217 & $<10^{-7}$ & LAGP.NN ($n$=50) & 4.773 & 0.7860 & 17.2 \\
		LAGP.NN ($n$=50) & 0.06229 & & LAGP.ALC ($n$=50) & 4.772 & & 55.7 \\
	\end{tabular}
	
	\caption{{\em Top:} RMSE (left) and score (right) vs.~log compute time over 30
		MC repetitions for the SIR experiment. Median statistics are denoted with bold
		markers. {\em Bottom:} Median RMSEs (left) and scores (right) in ascending
		rank. The $p$-values are for one-sided paired Wilcoxon tests between the model and the
		model directly below. Median compute time (in seconds) for each model is
		listed in the last column.
		\label{fig:sir_results}}
\end{figure*}

In this experiment we keep the same $\bar{N}=10000$ training set size as
Section \ref{ss:syn_results}, but now use a fixed degree of 10 replicates at
each location, a more common default design setup in the absence of external
information about regions of high/low variability.
We also keep a large $N'=10000$ testing set, which helps manage MC error in our out-of-sample metrics in this heteroskedastic
setting. The experiment's results are displayed in Figure \ref{fig:sir_results}.  The left
panel's vertical axis shows the RMSE values between each method's mean
predictions and the noisy testing observations. Variability across repetitions
in the data generating mechanism looms large compared to the differences in
RMSE values among the methods. Although RMSE/scores may look similar
marginally, within each MC repetition there is a clearer ordering. To expose
that, the left section of Figure \ref{fig:sir_results}'s table ranks the
methods based on their RMSE (lowest/best first). The reported $p$-values are
for one-sided paired Wilcoxon tests between a method's RMSE and the RMSE from the
next best fit. Observe, for example, that LIGP.qNorm (blue $+$'s) produces a
significantly lower distribution of RMSEs compared to the other models, with
LIGP.wIMSE (purple $\times$'s) and HetGP (black circles) not far behind.

Score (right panel) offers clearer distinctions between the methods, with
LIGP besting LAGP and HetGP via paired Wilcoxon test. Using score as a
metric highlights the differences in accuracy for the model estimates. By
weighting the squared error by the predictive variance, not all errors of the
same magnitude are considered equal (unlike in RMSE), which is crucial when
modeling a heteroskedastic process. LIGP.qNorm and
LIGP.wIMSE are statistically distinguishable from each other ($p$-value $< 0.01$)
and better than LAGP.NN with $n=100$ ($p$-value $< 10^{-9}$),
the next in the list. There is more variation in the computation times of the
LIGP models compared to LAGP, which is likely due to LIGP's tendency to need
more function evaluations for hyperparameter optimization. LIGP 
provides significant time savings despite an {\sf R} implementation.

\subsection{Ocean oxygen concentration model}
\label{ss:fksim}
Here we consider a stochastic simulator modeling oxygen
concentration deep in the ocean \citep{mckeague2005statistical}. This highly
heteroskedastic simulator uses MC to approximate the solution of an
advection-diffusion equation from computational fluid dynamics. There are four
inputs: latitude and longitude coordinates, and two diffusion coefficients. The
code required to run the simulation can be found in our repository.
Inputs are scaled to the unit cube $[0,1]^4$; the output is the ocean oxygen
concentration.

\begin{figure*}[ht!]
	\centering
	\begin{subfigure}{.49\textwidth}
		\includegraphics[trim=0 0 20 20 , clip,width=\textwidth]{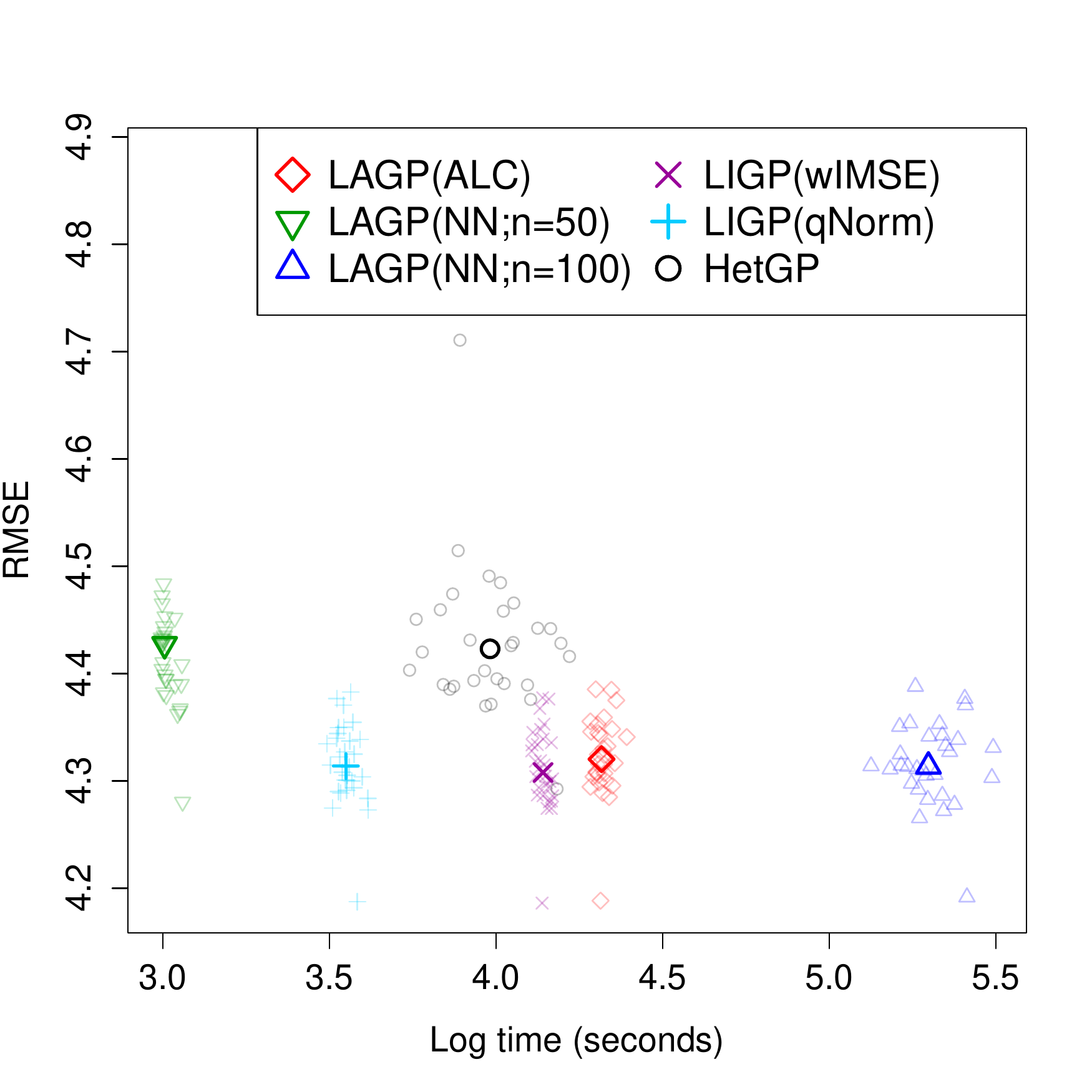}
	\end{subfigure}
	\begin{subfigure}{.49\textwidth}
		\includegraphics[trim=0 0 20 20 , clip,width=\textwidth]{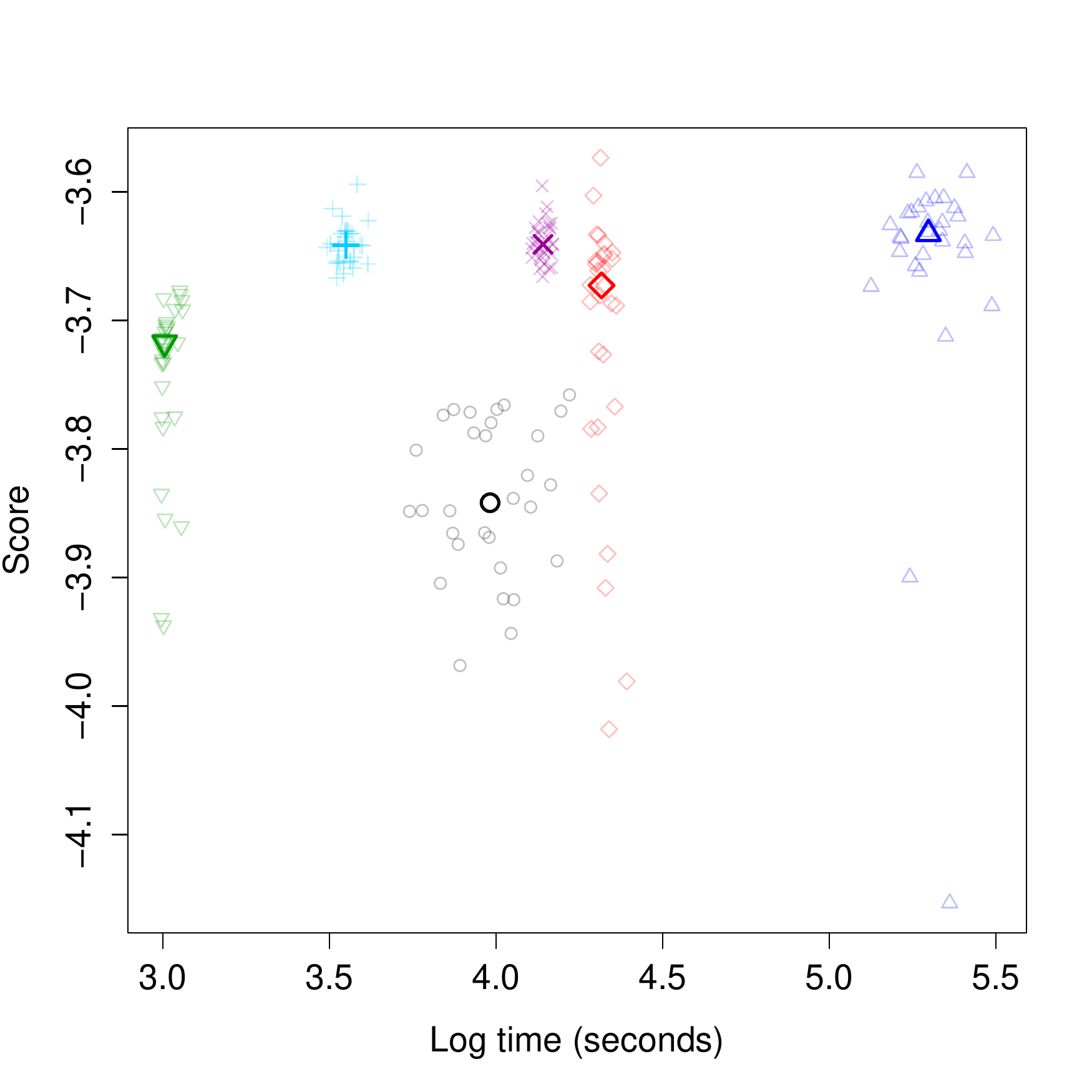}
	\end{subfigure}
	\vspace{0.25cm}
	\footnotesize
	\begin{tabular}{|l r r|| l r r r|}
		Model & RMSE & $p$-value & Model & Score & $p$-value & Time\\ \hline
		LIGP.wIMSE & 4.3078 & 0.0759 & LAGP.NN ($n$=100) & -3.633 & 0.0502 & 199.8 \\
		LAGP.NN ($n$=100) & 4.3135 & 0.0202 & LIGP.wIMSE & -3.641 & 0.0013 & 62.9 \\
		LIGP.qNorm & 4.3137 & 0.0005 & LIGP.qNorm & -3.642 & $<10^{-4}$ & 34.8 \\
		LAGP.ALC ($n$=50) & 4.3202 & $<10^{-9}$ & LAGP.ALC ($n$=50) & -3.673 & 0.0060 & 74.9 \\
		LAGP.NN ($n$=50) & 4.4271 & 0.9725 & LAGP.NN ($n$=50) & -3.717 & $<10^{-5}$ & 20.2 \\
		HetGP & 4.4230 & & HetGP & -3.842 & & 53.6\\
	\end{tabular}
	
	\caption{{\em Top:} RMSE (left) and score (right) vs.~log compute time over
		30 MC repetitions for Ocean Oxygen experiment. See Figure
		\ref{fig:sir_results} caption. \label{fig:fksim_results}}
\end{figure*}

We keep the same training/testing set sizes from the SIR experiment:
$\bar{N}=10000$ with 10 replicates each and $N'=10000$. Figure
\ref{fig:fksim_results} shows out-of-sample metrics similarly mirroring that
experiment. At first glance, HetGP (black circles) is clearly worse (higher
RMSEs, lower scores) than LIGP/LAGP. In 4d, using a subset of only 1000 unique locations does not provide enough
information to model this surface. The variation in compute time for
HetGP, due to challenges in modeling the latent noise, support this claim.

Focusing on LIGP and LAGP results, we find LIGP again among the best in RMSE,
score, and computation time. LAGP.NN with the larger neighborhood of $n=100$
(navy triangle) does well at modeling the mean and noise surfaces (its RMSE
and score medians are comparable to the LIGP models), but takes nearly 6 times
longer than LIGP.qNorm. HetGP (black circles) and LAGP.ALC (red diamonds) model the mean surface well,
producing slightly worse RMSEs, but struggle with UQ as measured by proper score.

\section{Bermudan Option Pricing} \label{sec:bermuda}

An application of high-throughput stochastic simulation arises in
computational methods for optimal stopping, which are particularly motivated
by quantitative finance contexts. Several types of financial contracts -- the
so-called American-type claims -- offer their buyer the option to exercise,
i.e.,~collect, her contract payoff at any time until its maturity. Pricing and
risk managing such a contract requires analysis of the respective optimal
exercise strategy, such deciding when to exercise the option.

Monte-Carlo-driven valuation of American-type claims consists of applying the
dynamic programming formulation to divide the global dynamic control problem
into local single-step optimization sub-problems. The latter revolve around
partitioning the input space (interpreted as the underlying prices at the
current time-step) into the \emph{exercise} and \emph{stopping regions}, to be
learned by constructing a certain statistical surrogate. More precisely, one
wishes to compare the $q$-value, namely the expected future payoff conditional
on not exercising now, with the immediate payoff. While the latter is a known
function,  the $q$-value is specified abstractly as a recursive conditional
expectation  and must be learned.

In the literature \citep{hetGP1,ludkovski2018kriging,ludkovski2020mlosp}, it
has been demonstrated that GP surrogates are well suited for this emulation
task, due to their flexibility, few tuning parameters, and
synergy with sequential design heuristics. Nevertheless, their cubic scaling
in design size $N$ is a serious limitation. The nature of the financial
context implies an input space is 2-5 dimensional
(1d problems are also common but are computationally straightforward)
and a very low signal-to-noise ratio. Consequently, Monte Carlo
methods for accurate estimation of the $q$-value call for $ N \gg 10^4$
simulations. 
\cite{ludkovski2018kriging} demonstrated that replication is pivotal to
addressing this scaling challenge; \cite{lyu2022adaptive} explored adaptive
replication. LIGP offers an additional boost by leveraging replication and
local approximation, simultaneously addressing speed, accuracy and
localization.

Relative to existing surrogates LIGP brings several improvements. First, it
overcomes limitations on the number of training simulations $N$. In previous
experiments working with matrices of size $\gg 2000$ was prohibitively slow.
This limited traditional GPs to data with dimension $d \le 3$ 
or excessive replication. Second, LIGP organically handles replicated designs
that are needed to separate signal from noise, which is heteroskedastic and
non-Gaussian in this application. Third, the localization intrinsic in LIGP is
beneficial to capture non-stationarity. Typically the $q$-value is flat at the
edges and has a strong curvature in the middle of the input space. Finally,
the control paradigm implies that the surrogate is intrinsically
prediction-driven -- the main use-case being to generate exercise rules along
a set of price scenarios. Thus, the surrogate is to be evaluated at a large
number of predictive locations and LIGP offers trivial parallelization that
vastly reduces compute time relative to plain GP variants.


\subsection{Illustrating the Exercise Strategy}

We have added LIGP regressors as a new choice to the {\tt mlOSP} library
\citep{ludkovski2020mlosp} for {\sf R}, which contains a suite of test cases
that focus on valuation of American-type options on multiple assets. This
allows benchmarking the LIGP module against alternatives, including plain
GP and {\tt hetGP} solvers that are already part of {\tt mlOSP}.

To fix ideas, we consider the $d$-dimensional Bermudan Max-Call with payoff function
$$
h_\mathrm{MaxCall}(k,\mathbf{x}) = e^{-r k\Delta t}( \max_{i \le d} x_i -
{\cal K})_+.$$ The parameter ${\cal K}$ is known as the strike, $r$ is the
continuously compounded interest rate and $\Delta t$ is the exercise frequency
measured in calendar time (e.g. daily, $\Delta t = 1/365$). The input space
$\mathbf{x} \in \mathbb{R}^d_+$ represents prices $x_1, \ldots, x_d$ of $d$
stocks. Thus, the buyer of the Max-Call is able to collect
$h_\mathrm{MaxCall}(k,\mathbf{X}(k\Delta t))$ dollars, the difference between
the largest stock price and the strike, at any one of the $K$ exercise
opportunities, $k=1,\ldots, K$. The choice of when to do so leads to $K$
decisions regarding whether to exercise the option at time step $k$ (assuming
it was not exercised yet) or not. Thus, we need $K-1$ surrogates for the
$q$-value $q(k,\x)$, $k=0,\ldots,K-1$.

The underlying stochastic simulator is based on generating trajectories of the
asset prices.  We stick to independent log-normal (Geometric Brownian motion)
dynamics for the price of the $i^{\mathrm{th}}$ asset,
\begin{align}\label{eq:gbm}
	dX_i(t) = (r - \delta_i) X_i(t) dt + \sigma_i X_i(t) dW_i(t), \qquad i=1,\ldots, d,
\end{align}
where $W_i(t)$ are independent standard Brownian motions, $\sigma_i$ is the
volatility, and $\delta_i$ the continuous dividend yield of the $i$-th asset.
This implies that given $X_i(s)$ for any $t>s$, $X_i(t)$ has a log-normal
distribution. Denote by $\mathbf{X}_{k:K} = \mathbf{X}_{k\Delta t}, \ldots,
\mathbf{X}_{K \Delta t}$ a realized trajectory of the $d$ asset prices on the
interval $[k\Delta t, K\Delta t]$.

The stochastic simulator yields empirical samples of $H_k(\mathbf{X}_k,
\mathbf{X}_{k+1}, \ldots, \mathbf{X}_K)$ where the aggregate function $H_k$,
representing the pathwise payoff, is a selector $$H_k(\x_k,\ldots,\x_K) =
\left\{\sum_{s=k+1}^{K} h(s, \mathbf{x}_{s}) 1_{A_s}\right\}-h(k, \x_k),$$ and
$A_k$ records the first pathwise exercise time along the given trajectory. The
objective of learning the $q$-value at step $k$ is equivalent to regressing
$H_k$ against $\x_k$ (exploiting the Markov property of the state dynamics) to
obtain the so-called \emph{timing value} $\hat{T}_k(\x_k)$ and identify
regions where the conditional mean is positive/negative.  Note that this task
must be done for \emph{each} time step $k=K-1,\ldots, 1$, and by construction
the outputs at step $k$ depend on the $\hat{T}_{k+1:K-1}$'s fitted previously.
This iterative estimation leads to non-trivial error back-propagation along
the financial time dimension in $k$.

Returning to the LIGP implementation, we have a sequence of surrogates that
must be constructed. For each surrogate, we are given a data set of size $N$
with $\bar{N}$ unique inputs in $\mathbb{R}^d_+$ and a constant number of
replicates $a$, $N = \bar{N} \cdot a$. For the experimental design we take an
LHS on a user-specified sub-domain. After fitting the surrogate for
$\hat{T}_k$, we need to predict on $N'=N$ unique outputs, which lie in the
training sub-domain but are otherwise distinct from the training inputs.

\begin{figure}[ht!]
	\centering
	\includegraphics[trim=0 0 0 20 , clip,width=.49\textwidth]{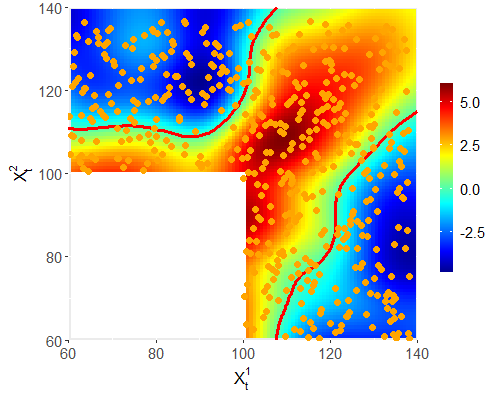}
	\includegraphics[trim=0 0 0 20 , clip,width=.49\textwidth]{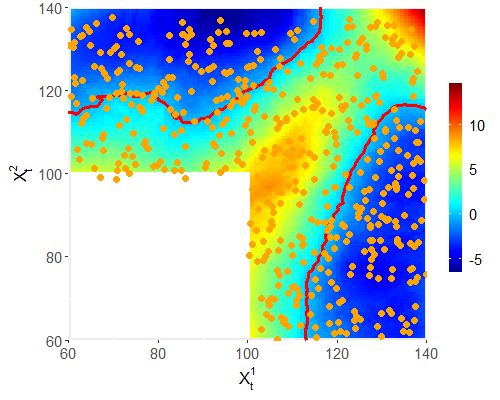}
	\caption{Fitted timing value and the corresponding exercise boundary (solid
		red curves) for the global GP (\emph{left}) and LIGP (\emph{right}) for the
		2-d Max-Call problem at time step $k=6$. The colors indicate the fitted
		$\hat{T}_k(x_1, x_2)$ and the points the training inputs $\x=(x_1,x_2)$. The
		bottom left corner is excluded from the input space since it has zero payoff.}
	\label{fig:bermudan2d}
\end{figure}

To illustrate how that looks,  Figure \ref{fig:bermudan2d} compares
global GP and LIGP fits for the timing value of the 2d Bermudan Max-Call
option. The model parameters are ${\cal K}=100$, $r=0.05$,
$\delta_1=\delta_2=0.1$, $\sigma_{1}=\sigma_{2} =0.2$, $T=3$ and $\Delta t =
1/3$ so that there are $K=9$ time periods. We display the fitted surrogates at
one intermediate time step, $k=6$.  In this problem, the option is
out-of-the-money when both $x_1$ and $x_2$ are below the strike ${\cal
	K}=100$, hence the exercise strategy is to continue in that region
and the latter (the white square in the bottom-left of each panel) is excluded
from the surrogate prediction. For both problems, we have a randomized
training set of $\bar{N} \approx 650$ inputs, with $a=25$ replicates each,
created by a LHS design. The LIGP surrogate uses $m=10,n=50$, with estimated
nugget and fixed lengthscale $\theta =1$.

The estimated strategy is to continue when the timing value $\hat{T}_k$ (color
coded in both panels) is positive and to stop when it is negative, with the
zero-contour (in red) demarcating the boundary of the stopping region. The
stopping region consists of two disconnected components, in the bottom-right
and upper-left; due to the symmetric choice of parameters, those regions (and
the entire response) are symmetric along the $x_1=x_2$ diagonal. Thus, the
response surface features a diagonal ridge with a global maximum around
$(100,100)$ and two basins on either side. The height of the ridge and the
depth of the basins change in $k$. The respective simulation variance is low
in regions where $T_k(\x) \simeq 0$ and is high in regions where $|T_k(\x)|$
is large.


We emphasize that Figure \ref{fig:bermudan2d} is purely diagnostic; actual
performance of the surrogate is based on the resulting estimated option
price. The latter is obtained by generating a test set of $N'$ scenarios and
using the collection of $K$ surrogates $\hat{T}_k$, $k=0,\ldots, K-1$ to
evaluate the resulting exercise times and ultimately the \emph{average}
payoff. Fixing a test set, the surrogate that yields a higher expected payoff
is deemed better. Traditionally, a test set is obtained by fixing an initial
$\mathbf{X}(0)$ and sampling $N'$ i.i.d.~paths of $\mathbf{X}_{0:K}$. For
example, taking $\mathbf{X}(0) = (90,90)$ and using the surrogates displayed
in Figure \ref{fig:bermudan2d} we obtain an estimated Max-Call option value
of 8.02 via LIGP and 8.00 via the global GP.
Note that this financial assessment of surrogate accuracy is not based on
standard IMSE-like metrics, but is driven by the accuracy of correctly
learning the exercise strategy, namely  the zero-contour of $H_k$. Hence, the
goal is to correctly specify the \emph{sign} of $T_k(\x)$ which determines
whether the option is exercised in the given state or not. Consequently,
errors in the magnitude of $\hat{T}_k$ are tolerated, while errors in the
sign of $\hat{T}_k$ will lead to degraded performance. 

\subsection{Results for Bermudan Option Pricing}

The 2d example was primarily for illustrative purposes; for a more challenging
case we take a 5d asymmetric Max-Call. We keep $T=3, \Delta t = 1/3, {\cal
	K}=100, r=0.05$ but adjust the parameters of the log-normal asset dynamics
\eqref{eq:gbm} to be $\delta_1 = \ldots =\delta_5 = 0.1$ and $\sigma_i = 0.08
\cdot i$ to have different volatilities. The setting therefore is no longer
symmetric and, due to larger variance, the third-through-fifth coordinates are
progressively more important. The initial starting point is taken to be
$\mathbf{X}(0) = (70,70,70,70,70)$.

We benchmark the results on a fixed test set of $N' = 25000$ paths of
$\mathbf{X}(t)$, comparing the resulting option price estimates.  Our
comparators include a global GP model {\tt homGP} and a {\tt hetGP} surrogate.
Note that the above are not pre-scaled and take inputs as-is. We train the
surrogates with an experimental design sampled from the log-normal density of
$\mathbf{X}(t)$, with $\bar{N}=2000$ unique inputs and $a=10$ replicates each.
In this case study we found that a relatively small number of inducing points
$m=30$ works well along with a  neighborhood size of $n=50$. We also left out various
embellishments that can further fine-tune each surrogate, aiming for a level
playing field.

\begin{figure}[ht!]
	\centering
	
	\includegraphics[trim=0 0 0 0 , clip,width=.6\textwidth]{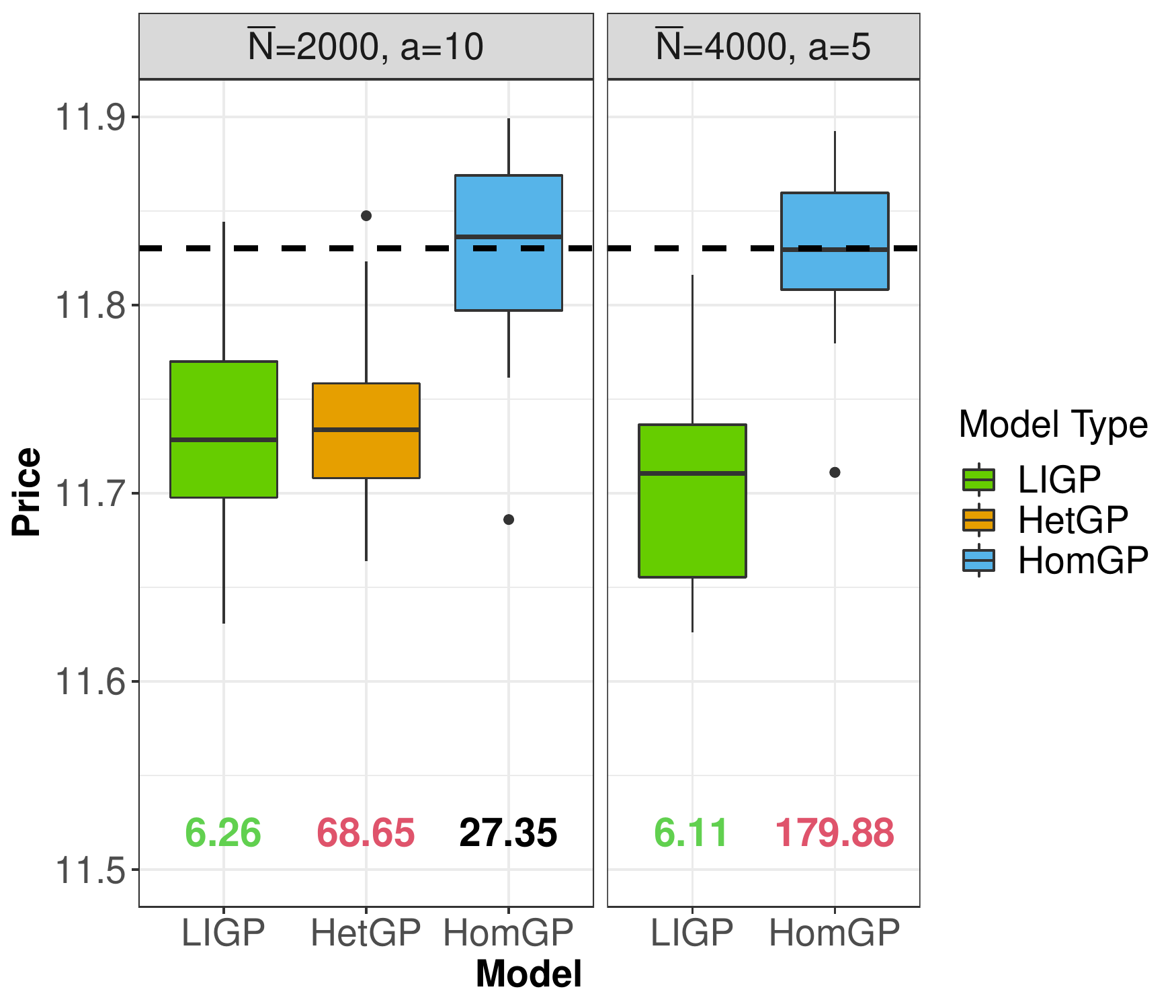}	
	
	\caption{Results for the 5d Asymmetric Bermudan
		Max-Call. Distributions of price over 30
		MC repetitions. Average compute time (mins) is below each box plot. All LIGP models (green) use the \texttt{qNorm} template. The
		default setting is $\bar{N}=2000,a=10,n=50$.}
	\label{tbl:bermudan5d}
\end{figure}

%



Figure~\ref{tbl:bermudan5d} presents the results, with the average compute times (in minutes) below each box plot. The reported compute time is
for the entire option pricing problem, i.e., across $K=9$ surrogates. The
reference option price is $11.83$ (black dashed line).
With a design of $\bar{N}=2000$, all three approaches yield
similar results that are close to optimal, but the compute times vary
substantially. LIGP is more than four times faster than {\tt homGP} and an order of
magnitude faster than {\tt hetGP}. In the problem context, methods that take
more than a handful of minutes are impractical, and these major speedup gains greatly outweigh the
small accuracy loss with LIGP. Not shown in
Figure~\ref{tbl:bermudan5d} are the results from {\tt laGP}. Using LAGP.ALC with $n=50$
is prohibitively slow in this example (over an hour, like {\tt hetGP}), while
using LAGP.NN yields very poor prices (below 11.5). Thus,
{\tt laGP} fails either in terms of speed or in terms of accuracy and is not a viable
alternative.


Exploration
of the input space tends to be the main driver of accuracy, so we consider
upgrading to $\bar{N}=4000$ and $a=5$ replicates. This keeps the same budget for
total number of simulations but is more exploratory. The new choice has minimal impact on the
compute time of LIGP that is driven by $\bar{N} \times a$ and is to zeroth-order independent of $\bar{N}$. Indeed,
the average compute times between $4000 \times 5$ (6.11 mins) and $2000 \times 10$ (6.26) are within 3\%
of each other. On the other hand, $\bar{N} \gg 10^3$ is too large for a global
GP: {\tt homGP} takes nearly three hours with $\bar{N}=4000$ and {\tt hetGP}
is not considered as it takes even longer.

Running times that exceed an hour (shown in red in Figure
\ref{tbl:bermudan5d}) are prohibitive.  A take-away here is that with a global
approach one is restricted in the choice of $\bar{N}$, while LIGP has
additional flexibility to survey the exploration-exploitation tradeoff. In
this particular setting, having more unique inputs does not yield substantial
performance improvement, but in related contexts it is a crucial factor for
developing an accurate exercise strategy.

\section{Discussion}
\label{sec:discussion}

Modern computing resources and advances in MC and agent-based modeling
strategies have combined of late to produce large simulation campaigns for
complex phenomena. The ability to adequately model such large data sets is
limited, especially when the simulator is heteroskedastic. Methods for
modeling large data sets exist, such as sparse matrix approximation and
divide-and-conquer, but are easily fooled into interpreting noise as signal
and provide inadequate UQ in the presence of input-dependent noise.

Local approximations (LAGP) on local subsets of data (neighborhoods) perform
well on predictive accuracy for deterministic data, because small local
subsets yield a substantial computational advantage over competing methods.
Scaling up to larger neighborhoods in the face of common replication-based
design strategies for separating signal from noise is problematic in that
context; larger neighborhoods wipe out that computational advantage. LIGP
frees up resources through inducing points, and thus allow for increases in
neighborhood size without severe computational bottlenecks, but is developed
solely for deterministic data. HetGP was designed for heterogeneous noisy
simulations, and replicated responses in the experimental design, and
consequently makes thrifty use of sufficient statistics and provides excellent
UQ.  But it is not a local approximation, and so it does not scale well to the
largest simulation campaigns.

Here we provide upgrades to the LIGP framework, essentially by combining these
three modern methods.  We first redefine LIGP neighborhoods based on the
number of unique design locations. Applying the common Woodbury
identities allows us to take advantage of replication. This provides calculations
on the order of the number of inducing points and unique locations.
Consequently LIGP may entertain much larger neighborhoods compared to LAGP and
the original version of LIGP. With these larger neighborhoods, we present evidence that
LIGP is able to better, and more quickly, separate signal from noise.

The promising results in this paper provide direction for further areas of
study. While current work relies on NN to build local neighborhoods,
variance-based alternatives (e.g., {\tt alcray} in {\tt laGP},
\cite{gramacy2016speeding}) may enhance modeling the mean surface. Extending
the kernel support to other families such as Mat\'ern \citep{Stein2012} may
also prove useful. Our empirical work relied on default choices of $(m,n)$
without much exploration.  A better understanding of these hyperparameters
(possibly through Bayesian optimization of out-of-sample RMSE or score)
could yield better defaults for simple, immediate implementation.

It is worth reinforcing every aspect of our local GP fits is unique to each
local neighborhood, providing a discontinuous noise (and mean) surface.  This
is in contrast to global methods (e.g., HomGP, HetGP).  We find that the
computational and other model fidelity advantages of such a local approach
outweigh the superficial disadvantages of a sometimes ``choppy'' predictive
surface. That said, local-global approach where noise information is shared
across neighborhoods \citep{edwards2021precision} could help smooth the noise
across the input space.

\section*{Acknowledgments}
We would like to thank the journal editor and referees for their thorough
review of this paper. They provided valuable insights and suggestions, helping
improve the narrative and context of this work. The authors declare they have
no conflict of interest. DAC and RBG recognize support from the National
Science Foundation (NSF) Grant DMS-1821258. Ludkovski is partially supported
by the NSF grant DMS-1821240.  Many thanks to Andrew Cooper for help with proofreading.

\bibliographystyle{jasa}
\bibliography{articles}

\section*{Appendix} \label{app:tau2}

Here we provide more details for re-expressing $\hat{\tau}^{2(\bar{n},m)}$ in Eq.~\eqref{eq:nuhatR} as Eq.~\eqref{eq:tau2_avg}. Based on \cite{hetGP1}, we can re-write $\hat{\tau}^{2(\bar{n},m)}$ as
$$\hat{\tau}^{2(\bar{n},m)} =  N^{-1}\left(\Y_n^\top\Omega^{-1(m)}_n\Y_n-\bar{\Y}_{\bar{n}}^\top \Lambda^{(m)}_{\bar{n}}\bar{\Y}_{\bar{n}}+\bar{n}\grave{\tau}^{2(\bar{n},m)}\right)$$
where $\grave{\tau}^{2(\bar{n},m)}=\bar{n}^{-1}\bar{\Y}_{\bar{n}}^\top \grave{\boldsymbol {\Sigma}}_{\bar{n}}^{-1}\bar{\Y}_{\bar{n}}$
and ${\grave{\boldsymbol \Sigma}}_{\bar{n}}^{(m)}=\kk_{\bar{n}m}\K_m^{-1}\kk_{\bar{n}m}^\top+A_{\bar{n}}^{-1}\Omega^{(m)}_{\bar{n}}$.
Recall that $\Omega_{\bar{n}}^{(m)}=\Delta_{\bar{n}}^{(m)} + g\I_{\bar{n}}$, containing the diagonal correction term  $\Delta_{\bar{n}}^{(m)}=\text{Diag}\{\K_{\bar{n}}-\kk_{\bar{n}m}\K_m^{-1}\kk_{\bar{n}m}^\top \}$. $\grave{\boldsymbol{\Sigma}}_{\bar{n}}^{(m)}$ contains $\Delta_{\bar{n}}^{(m)}$ in the reweighting, treating it as part of the pure noise. A correct expression of $\hat{\tau}^{2(\bar{n},m)}$ with unique-$\bar{n}$ calculations and $\bar{\Y}_{\bar{n}}$ is
$$\bar{\tau}^{2(\bar{n},m)}=\bar{n}^{-1}\bar{\Y}_{\bar{n}}^\top \bar{\boldsymbol{\Sigma}}_{\bar{n}}^{-1}\bar{\Y}_{\bar{n}},
$$
with $$\bar{\boldsymbol{\Sigma}}_{\bar{n}}^{(m)}=\kk_{\bar{n}m}\K_m^{-1}\kk_{\bar{n}m}^\top+\Delta_{\bar{n}}^{(m)} +gA_{\bar{n}}^{-1}.$$

Now we seek to write $\grave{\tau}^{2(\bar{n},m)}$ in terms of $\bar{\tau}^{2(\bar{n},m)}$ by redefining $\grave{\boldsymbol {\Sigma}}_{\bar{n}}^{-1}$ in terms of $\bar{\boldsymbol {\Sigma}}_{\bar{n}}^{-1}$. By using the Woodbury identity for $(\BB+\C\D\E)^{-1}$ from \eqref{eq:woodbury}, we let $\BB = \bar{\boldsymbol{\Sigma}}_{\bar{n}}^{(m)}=\kk_{\bar{n}m}\K_m^{-1}\kk_{\bar{n}m}^\top+\Delta_{\bar{n}}^{(m)} +gA_{\bar{n}}^{-1}$, $\D=(A_{\bar{n}}^{-1}-\I_{\bar{n}})\Delta_{\bar{n}}^{(m)}$, and $\C=\E=\I_{\bar{n}}$. Then
\begin{align}
	\grave{\boldsymbol {\Sigma}}_{\bar{n}}^{-1}&=(\BB+\D)\nonumber \\
	&=\BB^{-1} - \BB^{-1}(\D^{-1} + \BB^{-1})^{-1}\BB^{-1} \nonumber \\
	&=\bar{\boldsymbol{\Sigma}}_{\bar{n}}^{-1(m)}-\bar{\boldsymbol{\Sigma}}_{\bar{n}}^{-1(m)}((A_{\bar{n}}^{-1}-\I_{\bar{n}})^{-1}\Delta_{\bar{n}}^{-1(m)}+\bar{\boldsymbol{\Sigma}}_{\bar{n}}^{-1(m)})^{-1}\bar{\boldsymbol{\Sigma}}_{\bar{n}}^{-1(m)} \nonumber
\end{align}
Therefore, it follows that
\begin{align}
	\grave{\tau}^{2(\bar{n},m)}&=\bar{n}^{-1}\bar{\Y}_{\bar{n}}^\top \grave{\boldsymbol {\Sigma}}_{\bar{n}}^{-1}\bar{\Y}_{\bar{n}}\nonumber\\
	&=\bar{n}^{-1}\bar{\Y}_{\bar{n}}^\top \left( \bar{\boldsymbol{\Sigma}}_{\bar{n}}^{-1(m)}-\bar{\boldsymbol{\Sigma}}_{\bar{n}}^{-1(m)}(\D^{-1}+\bar{\boldsymbol{\Sigma}}_{\bar{n}}^{-1(m)})^{-1}\bar{\boldsymbol{\Sigma}}_{\bar{n}}^{-1(m)}\right)\bar{\Y}_{\bar{n}}\nonumber\\
	&= \bar{n}^{-1}\left(\bar{n}\bar{\tau}^{2(\bar{n},m)}-\bar{\Y}_{\bar{n}}^\top \bar{\boldsymbol{\Sigma}}_{\bar{n}}^{-1(m)}((A_{\bar{n}}^{-1}-\I_{\bar{n}})^{-1}\Delta_{\bar{n}}^{-1(m)}+\bar{\boldsymbol{\Sigma}}_{\bar{n}}^{-1(m)})^{-1}\bar{\boldsymbol{\Sigma}}_{\bar{n}}^{-1(m)}\bar{\Y}_{\bar{n}}\right).\nonumber
\end{align}
In turn,
\begin{align}
	\hat{\tau}^{2(\bar{n},m)} &=  N^{-1}\left(\Y_n^\top\Omega^{-1(m)}_n\Y_n-\bar{\Y}_{\bar{n}}^\top \Lambda^{(m)}_{\bar{n}}\bar{\Y}_{\bar{n}}+\bar{n}\grave{\tau}^{2(\bar{n},m)}\right)\nonumber\\
	&=N^{-1}\biggl(\bar{n}\bar{\tau}^{2(\bar{n},m)} + \Y_n^\top\Omega^{-1(m)}_n\Y_n-\bar{\Y}_{\bar{n}}^\top \Lambda^{(m)}_{\bar{n}}\bar{\Y}_{\bar{n}}-\nonumber\\
	& \qquad\qquad \bar{\Y}_{\bar{n}}^\top \bar{\boldsymbol{\Sigma}}_{\bar{n}}^{-1(m)}((A_{\bar{n}}^{-1}-\I_{\bar{n}})^{-1}\Delta_{\bar{n}}^{-1(m)}+\bar{\boldsymbol{\Sigma}}_{\bar{n}}^{-1(m)})^{-1}\bar{\boldsymbol{\Sigma}}_{\bar{n}}^{-1(m)}\bar{\Y}_{\bar{n}}\biggr).\nonumber
\end{align}
\end{document}